\documentclass[sigplan,10pt]{acmart}
\acmConference[PL'22]{ACM SIGPLAN Conference on Programming Languages}{January 01--03, 2022}{New York, NY, USA}
\acmYear{2022}
\acmISBN{} 
\acmDOI{} 
\startPage{1}

\setcopyright{none}

\usepackage{booktabs}   
\usepackage{subcaption} 

\usepackage{kotex}
\usepackage{fancyvrb}
\usepackage{nameref}
\usepackage{hyperref}
\usepackage{xspace}
\usepackage{cleveref}
  \Crefformat{chapter}{\S#2#1#3}
  \Crefformat{section}{\S#2#1#3}
  \Crefformat{figure}{Fig. #2#1#3}
\usepackage{iris}
\usepackage{wrapfig}
\usepackage{lipsum}
\usepackage{fancyvrb}
\usepackage{balance}
\usepackage{mathpartir}
\usepackage{stmaryrd}
\usepackage{cprotect}
\usepackage{pbox}
\usepackage[normalem]{ulem}
\usepackage{dashbox}
\usepackage{soul}
\usepackage{xcolor}
\usepackage{coqdoc}
\usepackage{multirow}
\usepackage{framed}
\usepackage{wasysym}
\usepackage{tabu}
\usepackage{enumitem}
\usepackage[super]{nth}
\definecolor{varpurple}{rgb}{0.4,0,0.4}
\definecolor{constrmaroon}{rgb}{0.6,0,0}
\definecolor{defgreen}{rgb}{0,0.6,0.1}
\definecolor{indblue}{rgb}{0,0,0.8}
\definecolor{kwred}{rgb}{0.8,0.1,0.1}
\definecolor{yjpurple}{rgb}{0.25,0,0.25}
\definecolor{yjgreen}{rgb}{0.25,0.5,0.25}

\def\coqdocvarcolor{yjpurple}
\def\coqdockwcolor{yjgreen}
\def\coqdoccstcolor{defgreen}
\def\coqdocindcolor{indblue}
\def\coqdocconstrcolor{constrmaroon}
\def\coqdocmodcolor{defgreen}
\def\coqdocaxcolor{varpurple}
\def\coqdoctaccolor{black}

\def\coqdockw#1{{\color{\coqdockwcolor}{{\texttt{#1}}}}}
\def\coqdocvar#1{{\color{\coqdocvarcolor}{\hspace{1pt}\texttt{#1}\hspace{1pt}}}}
\def\coqdoccst#1{{\color{\coqdoccstcolor}{\textrm{#1}}}}
\def\coqdocind#1{{\color{\coqdocindcolor}{\textsf{#1}}}}
\def\coqdocconstr#1{{\color{\coqdocconstrcolor}{\textsf{#1}}}}
\def\coqdocmod#1{{{\color{\coqdocmodcolor}{\textsc{\textsf{#1}}}}}}
\def\coqdocax#1{{{\color{\coqdocaxcolor}{\textsl{\textrm{#1}}}}}}

\begin{document}

\sethlcolor{lightgray}

\title[Conditional Contextual Refinement]{Conditional Contextual Refinement (CCR)
}         


\author{Youngju Song}
\affiliation{
  \institution{Seoul National University}            
  \country{Korea}                    
}
\email{youngju.song@sf.snu.ac.kr}          

\author{Minki Cho}
\affiliation{
  \institution{Seoul National University}            
  \country{Korea}                    
}
\email{minki.cho@sf.snu.ac.kr}         

\author{Dongjae Lee}
\affiliation{
  \institution{Seoul National University}            
  \country{Korea}                    
}
\email{dongjae.lee@sf.snu.ac.kr}         

\author{Chung-Kil Hur}
\affiliation{
  \institution{Seoul National University}            
  \country{Korea}                    
}
\email{gil.hur@sf.snu.ac.kr}         

\newcommand{\ie}{\textit{i.e.,} }
\newcommand{\cf}{\textit{cf.} }
\newcommand{\eg}{\textit{e.g.,} }
\newcommand{\etal}{\textit{et al.}}

\newcommand{\myhspace}{\hspace*{-.4pt}}

\newcommand{\gil}[1]{\textcolor{red}{\textbf{gil: #1}}}
\newcommand{\yj}[1]{\textcolor{red}{\textbf{yj: #1}}}
\newcommand{\yjj}[1]{\textcolor{red}{#1}}
\newcommand{\minki}[1]{\textcolor{green!60!black}{\textbf{minki: #1}}}
\newcommand{\ldj}[1]{\textcolor{green!60!black}{\textbf{ldj: #1}}}
\newcommand{\todo}[1]{\textcolor{red}{\textbf{TODO: #1}}}


\newcommand{\code}[1]{\texttt{#1}\xspace}
\newcommand{\valr}{r}
\newcommand{\beh}[1]{\text{Beh}({#1})}
\newcommand{\behems}{\code{$\overline{\text{Beh}}$}}
\newcommand{\crems}{AL\xspace}
\newcommand{\CREMS}{abstraction logic\xspace}
\newcommand{\ems}{\textrm{EMS}\xspace}
\newcommand{\imp}{IMP\xspace}
\newcommand{\spc}{SPC\xspace}
\newcommand{\gen}{\textrm{\spc}\xspace}
\newcommand{\CR}{CR}
\newcommand{\checker}{SpecChecker\xspace}
\newcommand{\barr}{\noindent\makebox[\linewidth]{\rule{\textwidth}{0.4pt}}}
\newcommand{\exitval}{0}
\newcommand{\asm}[1]{\textcolor{purple}{#1}}
\newcommand{\grt}[1]{\textcolor{blue}{#1}}
\newcommand{\Grt}[1]{\grt{\text{Grt(#1)}}}
\newcommand{\Asm}[1]{\asm{\text{Asm(#1)}}}
\newcommand{\AsmL}{\asm{\text{Asm}}}
\newcommand{\GrtL}{\grt{\text{Grt}}}

\newcommand{\link}{\circ}
\newcommand{\iris}{Iris\xspace}
\newcommand{\vst}{VST\xspace}
\newcommand{\lectx}{\le_{\text{ctx}}}
\newcommand{\leclosed}{\textcolor{green!60!black}{\le_{\text{NOT CTX REFINEMENT FIXME}}}}
\newcommand{\boilerplate}[1]{\textcolor{white!60!black}{#1}}
\newcommand{\mires}{\sigma}

\newcommand{\originaltxt}[1]{}
\newcommand{\hide}[1]{}
\newcommand{\unhide}[1]{#1}

\newcommand{\myparagraph}[1]{\paragraph{\hspace*{-3.8mm}\bfseries{#1}}}
\definecolor{grey}{rgb}{0.5,0.5,0.5}
\newcommand{\caselabel}[1]{\begin{small}{\color{grey}{(\textsc{#1})}}\end{small}}
\newcommand{\mycomment}[1]{\begin{small}{\vspace{1.5mm}\color{grey}{(*****\;#1\;*****)}}\end{small}}

\newcommand{\twoA}{{2\textrm{A}}}
\newcommand{\twoB}{{2\textrm{B}}}

\newenvironment{stackAux}[2]{%
  \setlength{\arraycolsep}{0pt}
  \begin{array}[#1]{#2}}{
  \end{array}}
\newenvironment{stackCC}{
  \begin{stackAux}{c}{c}}{\end{stackAux}}
\newenvironment{stackCL}{
  \begin{stackAux}{c}{l}}{\end{stackAux}}
\newenvironment{stackTL}{
  \begin{stackAux}{t}{l}}{\end{stackAux}}
\newenvironment{stackTR}{
  \begin{stackAux}{t}{r}}{\end{stackAux}}
\newenvironment{stackBC}{
  \begin{stackAux}{b}{c}}{\end{stackAux}}
\newenvironment{stackBL}{
  \begin{stackAux}{b}{l}}{\end{stackAux}}

\NewDocumentCommand \ahoare {m m m O{}}{
	\curlybracket{#1}& \spac #2 \spac &\curlybracket{#3}_{#4}%
}

\newcommand{\impl}{{\textsl{impl}}}
\newcommand{\spec}{{\textsl{spec}}}
\newcommand{\inimpl}[1]{\ensuremath{#1_\impl}}
\newcommand{\inspec}[1]{\ensuremath{#1_\spec}}

\newcommand{\cfbox}[2]{%
    \colorlet{currentcolor}{.}%
    {\color{#1}%
    \fbox{\color{currentcolor}#2}}%
}
\newcommand{\cdashbox}[2]{%
    \colorlet{currentcolor}{.}%
    {\color{#1}%
    \dashbox{\color{currentcolor}#2}}%
}
\newcommand{\defeq}{\ensuremath{\stackrel{\text{def}}{=}}}
\newcommand{\defcoind}{\ensuremath{\stackrel{\text{coind}}{=}}}

\newcommand{\mwimp}[1]{\text{$P_\text{\code{MW}}^{#1}$}\xspace}
\newcommand{\mwmid}{\text{$I_\text{\code{MW}}$}\xspace}
\newcommand{\mwmidcond}{\text{$I_\text{\code{MW}}^\text{cond}$}\xspace}
\newcommand{\mwabs}{\text{$A_\text{\code{MW}}$}\xspace}
\newcommand{\mwabscond}{\text{$A_\text{\code{MW}}^\text{cond}$}\xspace}

\newcommand{\memimp}{\text{$P_\text{\code{Mem}}$}\xspace}
\newcommand{\memmid}{\text{$I_\text{\code{Mem}}$}\xspace}
\newcommand{\memmidcond}{\text{$I_\text{\code{Mem}}^\text{cond}$}\xspace}
\newcommand{\memabs}{\text{$A_\text{\code{Mem}}$}\xspace}
\newcommand{\memabscond}{\text{$A_\text{\code{Mem}}^\text{cond}$}\xspace}

\newcommand{\mapimp}{\text{$P_\text{\code{Map}}$}\xspace}
\newcommand{\mapmid}{\text{$I_\text{\code{Map}}$}\xspace}
\newcommand{\mapmidcond}{\text{$I_\text{\code{Map}}^\text{cond}$}\xspace}
\newcommand{\mapabs}{\text{$A_\text{\code{Map}}$}\xspace}
\newcommand{\mapabscond}{\text{$A_\text{\code{Map}}^\text{cond}$}\xspace}

\newcommand{\onceimp}{\text{$P_\text{\code{Once}}$}\xspace}
\newcommand{\onceabs}{\text{$A_\text{\code{Once}}$}\xspace}
\newcommand{\onceabstry}{\text{$A_\text{\code{Once}}^\text{try}$}\xspace}
\newcommand{\onceabscond}{\text{$A_\text{\code{Once}}^\text{cond}$}\xspace}
\newcommand{\testimp}[1]{\text{$P_\text{\Test{#1}}$}\xspace}
\newcommand{\testabs}[1]{\text{$A_\text{\Test{#1}}$}\xspace}
\newcommand{\testabscond}[1]{\text{$A_\text{\Test{#1}}^\text{cond}$}\xspace}
\newcommand{\appimp}{\text{$P_\text{\code{App}}$}\xspace}
\newcommand{\appmid}{\text{$I_\text{\code{App}}$}\xspace}
\newcommand{\appmidcond}{\text{$I_\text{\code{App}}^\text{cond}$}\xspace}
\newcommand{\appabs}{\text{$A_\text{\code{App}}$}\xspace}
\newcommand{\appabscond}{\text{$A_\text{\code{App}}^\text{cond}$}\xspace}

\newcommand{\echoimpl}{\text{$I_\text{\code{Echo}}$}\xspace}
\newcommand{\echospec}{\text{$S_\text{\code{Echo}}$}\xspace}
\newcommand{\echoabs}{\text{$A_\text{\code{Echo}}$}\xspace}
\newcommand{\echogen}{\text{$G_\text{\code{Echo}}$}\xspace}
\newcommand{\stackimpl}{\text{$I_\text{\code{Stack}}$}\xspace}
\newcommand{\stackspec}{\text{$S_\text{\code{Stack}}$}\xspace}
\newcommand{\stackabs}{\text{$A_\text{\code{Stack}}$}\xspace}
\newcommand{\stackgen}{\text{$G_\text{\code{Stack}}$}\xspace}
\newcommand{\memimpl}{\text{$I_\text{\code{Mem}}$}\xspace}
\newcommand{\memspec}{\text{$S_\text{\code{Mem}}$}\xspace}
\newcommand{\memtop}{\text{$\overline{A}_\text{\code{Mem}}$}\xspace}
\newcommand{\memgen}{\text{$G_\text{\code{Mem}}$}\xspace}
\newcommand{\bwimpl}{\text{$I_\text{\code{Bw}}$}\xspace}
\newcommand{\bwspec}{\text{$S_\text{\code{Bw}}$}\xspace}
\newcommand{\bwgen}{\text{$G_\text{\code{Bw}}$}\xspace}

\newcommand{\sys}[1]{{\code{\it{#1}}}}
\newcommand{\sysprint}{\sys{print}}
\newcommand{\sysgetint}{\sys{getint}}

\newcommand{\kw}[1]{{\code{\textbf{#1}}}}
\newcommand{\kwreturn}{\kw{return}}
\newcommand{\kwipc}{\kw{apc}}
\newcommand{\kwret}{\kw{ret}}
\newcommand{\kwdef}{\kw{def}}
\newcommand{\kwspec}{\kw{spec}}
\newcommand{\kwfriend}{\kw{friend}}
\newcommand{\kwcontext}{\kw{context}}
\newcommand{\kwvar}{\kw{var}}
\newcommand{\kwif}{\kw{if}}
\newcommand{\kwthen}{\kw{then}}
\newcommand{\kwelse}{\kw{else}}
\newcommand{\kwmatch}{\kw{match}}
\newcommand{\kwwith}{\kw{with}}
\newcommand{\kwend}{\kw{end}}
\newcommand{\kwassume}{{\color{purple}\kw{assume}}}
\newcommand{\kwguarantee}{{\color{blue}\kw{guarantee}}}
\newcommand{\kwgrnt}{{\color{blue}\kw{grnt}}}
\newcommand{\ASSUME}{\textcolor{purple}{\code{ASSUME}}}
\newcommand{\GUARANTEE}{\textcolor{blue}{\code{GUARANTEE}}}
\definecolor{darkblue}{rgb}{0.0, 0.0, 0.55}
\definecolor{darkred}{rgb}{0.55, 0.0, 0.0}
\newcommand{\kwupdate}{\ensuremath{\rightsquigarrow=}}
\newcommand{\kwaddp}{\kw{\hspace{0.3mm}+=}}
\newcommand{\kwsubp}{\kw{\hspace{0.3mm}-=}}
\newcommand{\kwmodule}{\kw{Module}}
\newcommand{\kwdefault}{\kw{default}}
\newcommand{\kwskip}{\kw{skip}}
\newcommand{\kwfpu}{\kw{fpu}}
\newcommand{\kwminus}{\kw{minus}}

\newcommand{\kwchoose}{\kw{choose}}
\newcommand{\kwtake}{\kw{take}}
\newcommand{\kwsyscall}{\kw{syscall}}
\newcommand{\kwloop}{\kw{loop}}
\newcommand{\kwbreak}{\kw{break}}
\newcommand{\kwcall}{\kw{call}}
\newcommand{\kwlocal}{\kw{local}}
\newcommand{\kwub}{\kw{UB}}
\newcommand{\kwnb}{\kw{NB}}
\newcommand{\kwfun}{\kw{fun}}
\newcommand{\kwwhile}{\kw{while}}

\newcommand{\res}{\mbox{$\sigma$}}
\newcommand{\mres}{\boilerplate{\code{mres}}}
\newcommand{\mresb}{\code{mres}}
\newcommand{\eres}{\code{eres}}
\newcommand{\lres}{\code{lres}}
\newcommand{\frm}{\code{frm}}
\newcommand{\ms}{\code{ms}}

\newcommand{\resframe}{\code{res\_frame}}
\newcommand{\kwcanonpcm}{\code{$\textrm{PCM}_\Cannon$}}

\newcommand{\kwmalloc}{\kw{malloc}}
\newcommand{\kwfree}{\kw{free}}
\newcommand{\kwload}{\kw{load}}
\newcommand{\kwstore}{\kw{store}}
\newcommand{\kwrepeat}{\kw{repeat}}

\newcommand{\kwpure}{\textrm{TRIVIAL}}

\newcommand{\Snone}{S_{*}}
\newcommand{\snone}{s_{*}}

\newcommand{\optionm}{\ensuremath{\textdom{Option}}}

\newcommand{\fsem}{f_\textrm{sem}}
\newcommand{\Sf}{S_\code{f}}
\newcommand{\RP}{\code{RP}}
\newcommand{\SC}{\code{SC}}
\newcommand{\AD}{\code{AD}}
\newcommand{\IO}{\code{IO}}
\newcommand{\MW}{\code{MW}}
\newcommand{\Once}{\code{Once}}
\newcommand{\Test}[1]{\mbox{$\code{Test}_{#1}$}}
\newcommand{\App}{\code{App}}
\newcommand{\Map}{\code{Map}}
\newcommand{\Echo}{\code{Echo}}
\newcommand{\Mem}{\code{Mem}}
\newcommand{\Stack}{\code{Stack}}
\newcommand{\Main}{\code{Main}}
\newcommand{\main}{\code{main}}
\newcommand{\MF}{\code{F}}
\newcommand{\mf}{\code{f}}
\newcommand{\Cannon}{\code{Cannon}}
\newcommand{\fire}{\code{fire}}
\newcommand{\kwundef}{\code{Undef}}
\newcommand{\kwemp}{\code{$\munit$}}
\newcommand{\ball}{\code{Ball}}
\newcommand{\slball}{\ownGhost{}{\code{Ball}}}
\newcommand{\ready}{\code{Ready}}
\newcommand{\fired}{\code{Fired}} 
\newcommand{\numfire}{\code{NUM\_FIRE}}
\newcommand{\powder}{\code{powder}}

\newcommand{\Do}{\code{Do}}
\newcommand{\Init}{\code{Init}}
\newcommand{\Run}{\code{Run}}
\newcommand{\Both}{\code{Both}}
\newcommand{\MWis}{\code{MWhas}}
\newcommand{\contains}{\!\!\ensuremath{\mapsto_{\Map}}\!}
\newcommand{\pointsto}{\ensuremath{\mapsto}\!}

\newcommand{\madd}{+}
\newcommand{\nullptr}{\code{NULL}}

\newcommand{\usable}[1]{\ownGhost{}{\textrm{Expect(#1)}}}
\newcommand{\usablex}[1]{\textrm{Expect(#1)}}
\newcommand{\usablexnoarg}[1]{\textrm{Expect}}
\newcommand{\being}[1]{\ownGhost{}{\textrm{Indeed(#1)}}}
\newcommand{\beingx}[1]{\textrm{Indeed(#1)}}
\newcommand{\beingxnoarg}{\textrm{Indeed}}
\newcommand{\full}[1]{\ownGhost{}{\textrm{Ok(#1)}}}
\newcommand{\fullx}[1]{\textrm{Ok(#1)}}
\newcommand{\fullxnoarg}{\textrm{Ok}}
\newcommand{\unitt}{\textrm{Emp}}
\newcommand{\wrong}{\textrm{Wrong}}
\newcommand{\true}{\textrm{T}}
\newcommand{\false}{\textrm{F}}
\newcommand{\colorcode}{\textrm{hex}}
\newcommand{\bb}{\texttt{b}}
\newcommand{\wdef}{\mval}

\newcommand{\prefire}{\code{pre\_fire}}
\newcommand{\postfire}{\code{post\_fire}}
\newcommand{\premain}{\code{pre\_main}}
\newcommand{\postmain}{\code{post\_main}}

\newcommand{\name}{\code{name}}

\newcommand{\upcast}[1]{{\ensuremath{{\uparrow}{#1}}}}
\newcommand{\downcastU}[1]{\code{\color{purple}$?_{#1}$}}
\newcommand{\downcastN}[1]{\code{\color{blue}$!_{#1}$}}
\newcommand{\unwrapU}{{\color{purple}?}}
\newcommand{\unwrapN}{{\color{blue}!}}
\newcommand{\empval}{\code{ev}}
\newcommand{\asgn}{\code{$\leftarrow$}}

\newcommand{\sljudge}[3]{{#1 \model #2 : #3}}
\newcommand{\spcmod}[3]{{[#1 \rtimes #2 : #3]}}
\newcommand{\safe}[1]{{#1 \text{ safe}}}
\newcommand{\Safe}{{\trm{Safe}\hspace*{.5pt}}}
\newcommand{\spcabs}[1]{{#1^\text{A}}}
\newcommand{\divcode}{\,\code{/}\,}
\newcommand{\modcode}{\,\code{\%}\,}
\newcommand{\prt}{$print\hspace*{-.5pt}$}
\newcommand{\cmt}[1]{\textcolor{white!60!black}{//#1}}

\newcommand{\mydashbox}[3]{\hspace*{1pt}\textcolor{#1}{\dashbox{\parbox{#2}{#3}}}}
\setlength{\dashdash}{1pt}
\setlength{\dashlength}{2pt}

\newcommand{\option}{\code{option}}
\newcommand{\vunit}{\code{()}}
\newcommand{\vlist}{\code{list}}
\newcommand{\lsthead}{\textrm{head}}
\newcommand{\lsttail}{\textrm{tail}}
\newcommand{\lstnz}{\textrm{nonzero}}
\newcommand{\lstlength}{\textrm{len}}
\newcommand{\some}{\code{Some}}
\newcommand{\none}{\code{None}}
\newcommand{\vval}{\code{val}}
\newcommand{\vptr}{\code{ptr}}
\newcommand{\vint}{\ensuremath{\code{int}_{64}}}
\newcommand{\vintopt}{\ensuremath{\code{int}_{64}^{?}}}
\newcommand{\vbool}{\code{bool}}
\newcommand{\vundef}{\code{vundef}}
\newcommand{\rplift}[1]{\lc #1 \rc}
\newcommand{\lc}{\ulcorner}
\newcommand{\rc}{\urcorner}
\newcommand{\own}[1]{\ownGhost{}{#1}}
\newcommand{\opure}[1]{\code{$\langle{#1}\rangle$}}
\newcommand{\otop}{\infty}
\newcommand{\lrepeat}{\textrm{repeat}}
\newcommand{\blk}{\textrm{b}}
\newcommand{\ofs}{\textrm{ofs}}
\newcommand{\xp}{\textrm{p}}
\newcommand{\xv}{\textrm{v}}
\newcommand{\xx}{\textrm{x}}
\newcommand{\xr}{\textrm{r}}
\newcommand{\handle}{h}
\newcommand{\stack}{\textrm{stk}}
\newcommand{\stackz}{\textrm{estk}}

\newcommand{\argp}{x}
\newcommand{\argv}{x_\textrm{a}}
\newcommand{\argo}{d}
\newcommand{\retp}{r}
\newcommand{\retv}{r_\textrm{a}}
\newcommand{\lst}{\ell}

\newcommand{\tyany}{\anyty}
\newcommand{\tyt}{\textrm{T}}
\newcommand{\mypoint}[1]{\textbf{\emph{(#1)}}}
\newcommand{\isstack}[2]{{\textrm{is\_stk}\;#1\;#2}}
\newcommand{\isbag}[2]{{\textrm{is\_bag}\;#1\;#2}}
\newcommand{\isestack}[2]{{\textrm{is\_estk}\;#1\;#2}}
\newcommand{\rom}[1]{\uppercase\expandafter{\romannumeral #1\relax}}
\newcommand{\myset}[2]{{#1|_{#2}}}

\newcommand{\llink}{\link}
\newcommand{\llinkall}{\ensuremath{\fullmoon}}
\newcommand{\plink}{\ensuremath{\bullet}}
\newcommand{\trm}[1]{\textrm{#1}}

\newcommand{\lland}{\,\land\,}
\newcommand{\emsmod}{\trm{Mod}}
\newcommand{\emsmods}{\trm{Mods}}
\newcommand{\LD}{\trm{LD}}
\newcommand{\mlift}{\trm{lift}}

\newcommand{\anyty}{\code{Any}}

\newcommand{\toabs}[1]{[#1]}
\newcommand{\setofz}[1]{\{#1\}}
\newcommand{\metaspec}[1]{\overline{#1}}

\newcommand{\kwgetcaller}{\kw{get\_caller}}

\newcommand{\world}{\ensuremath{\mathcal{W}}}
\newcommand{\simrel}{I}
\newcommand{\transA}{toAbs}
\newcommand{\transAS}{toAbspec}
\newcommand{\ASFun}{\trm{FunDef}}
\newcommand{\ASCall}{\trm{CallDef}}
\newcommand{\ASIPC}{\trm{APCDef}}
\newcommand{\ASIPCi}{\trm{APCiDef}}


\definecolor{cverbbg}{gray}{0.93}

\newenvironment{lcverbatim}
 {\SaveVerbatim{cverb}}
 {\endSaveVerbatim
  \flushleft\fboxrule=0pt\fboxsep=.5em
  \colorbox{cverbbg}{%
    \makebox[\dimexpr\linewidth-2\fboxsep][l]{\BUseVerbatim{cverb}}%
  }
  \endflushleft
}

\definecolor{shadecolor}{gray}{0.93}
\newcommand{\mbind}{\ensuremath{>\!\!>\!=}}
\newcommand{\mcat}{\ensuremath{>\!\!>\!\!>}}

\newcommand{\rProp}{\mathbf{rProp}}

\newcommand{\ord}{\code{ord}}

\newcommand{\Meta}{\code{W}}
\newcommand{\Dep}{\code{D}}
\newcommand{\Pre}{\code{P}}
\newcommand{\Post}{\code{Q}}
\newcommand{\meta}{w}
\newcommand{\cnd}{\code{c}}

\newcommand{\impcomp}[1]{{\llparenthesis #1 \rrparenthesis}}

\DeclareRobustCommand\model{\mathrel{|}\joinrel\mkern-.5mu\mathrel{\textrm{--}}}


\newcommand{\anyv}{Any}
\newcommand{\propb}{\mathbb{P}}
\newcommand{\typeb}{Type}

\newcommand{\effectarrow}{\hookrightarrow}

\newcommand{\itree}[2]{itree #1 #2}

\newcommand{\fname}{string}
\newcommand{\mname}{string} 

\newcommand{\chooseE}{\code{Choose}}
\newcommand{\takeE}{\code{Take}}
\newcommand{\syscallE}{\code{Obs}}
\newcommand{\ipcE}{\code{APC}}

\newcommand{\getNameE}{\code{GetCaller}}
\newcommand{\putE}{\code{Put}}
\newcommand{\getE}{\code{Get}}
\newcommand{\callE}{\code{Call}}

\newcommand{\EventE}{\ensuremath{\trm{E}_{\text{prim}}}}
\newcommand{\ModE}{\ensuremath{\trm{E}_{\text{mod}}}}
\newcommand{\EmsE}{\ensuremath{\trm{E}_{\text{\ems}}}}
\newcommand{\SpcE}{\ensuremath{\trm{E}_{\text{\gen}}}}

\newcommand{\unitset}{()}

\newcommand{\SLtriple}{SL\_triple}
\newcommand{\SLspec}{SL\_spec}
\newcommand{\SLspecs}{SL\_spectable}

\newcommand{\mfun}[1]{function#1}

\newcommand{\spcmodforget}[2]{|#1|$_{\text{#2}}$}

\newcommand{\spcfun}[4]{{\{#1 \rtimes #2 : #3 / #4\}}}
\newcommand{\spcfunforget}[3]{|#1 / #2|\texorpdfstring{\textsubscript{#3}}}

\newcommand{\SLfun}{SL\_decorate}

\newcommand{\SLpre}{SL\_arg}
\newcommand{\SLpost}{SL\_ret}
\newcommand{\SLcall}{SL\_call}
\newcommand{\SLbody}{SL\_body}
\newcommand{\SLIPC}{SL\_APC}
\newcommand{\SLsafe}{SL\_safe}

\newcommand{\rprop}{rProp}

\newcommand{\all}{g}
\newcommand{\names}{ns}

\newcommand{\simle}{\ensuremath{\sqsubseteq_{_{\world}}}}
\newcommand{\ievent}{effect}

\newcommand{\kwrp}{\code{res\_r}}
\newcommand{\kwap}{\code{res\_x}}

\newcommand{\iside}{\code{i}}
\newcommand{\aside}{\code{a}}

\newcommand{\myhrule}{\\[2mm]\hrule}
\newcommand{\mynewpage}{\newpage\noindent}

\newcommand{\fn}{\code{$f\!n$}}

\newcommand{\xorvp}{xr}
\newcommand{\xorvv}{xr_\trm{a}}

\newcommand{\simplera}{\exm\,(\{1\})}  
\newcommand{\appra}{\trm{AppPCM}}
\newcommand{\half}{\exinj(1)} 
\newcommand*{\ownGhostMap}[2]{(#2)}

\newcommand{\estep}[1]{\stackrel{#1}{\hookrightarrow}}


\begin{abstract}
  Contextual refinement (CR) is one of the standard notions of specifying open programs.
CR has two main advantages: $(i)$ (horizontal and vertical) compositionality that
allows us to decompose a large contextual refinement into many smaller ones enabling modular and incremental verification,
and $(ii)$ no restriction on programming features thereby allowing, \eg
mutually recursive, pointer-value passing, and higher-order functions.
However, CR has a downside that it cannot impose conditions on the context
since it quantifies over \emph{all} contexts, which indeed plays a key role in support of full compositionality and programming features.

In this paper, we address the problem of finding a notion of refinement that
satisfies all three requirements: support of full compositionality, full (sequential) programming features,
and rich conditions on the context.
As a solution, we propose a new theory of refinement, called CCR (Conditional Contextual Refinement), and
develop a verification framework based on it,
which allows us to modularly and incrementally verify a concrete module against an abstract module
under separation-logic-style pre and post conditions about external modules.
It is fully formalized in Coq and provides a proof mode that combines $(i)$ simulation reasoning about preservation of side effects such as IO events and termination
and $(ii)$ propositional reasoning about pre and post conditions.
Also, the verification results are combined with CompCert, so that we formally establish behavioral refinement from top-level abstract programs, all the way down to their assembly code.

\end{abstract}

\ccsdesc[500]{Software and its engineering~General programming languages}
\ccsdesc[300]{Social and professional topics~History of programming languages}


\maketitle

\section{Introduction}
\label{sec:introduction}

\begin{figure*}[t]
\begin{minipage}[t]{0.247\textwidth}
$\mwimp{1}$ :=
\begin{Verbatim}[commandchars=\\\{\},codes={\catcode`$=3},fontsize=\small,baselinestretch=0.97]
[\kwmodule \MW{}]
\color{defgreen}\kwlocal arr
\kwlocal map
\kwdef main() $\equiv$
  \color{defgreen}arr := \Mem.alloc(100)
  map := \Map.new()
  \App.init()
  \kwwhile (true)
    \App.run()
\kwdef put(i:\vint{},v:\vint{}) $\equiv$


  \color{defgreen}\kwif (0 <= i && i < 100)
    \color{defgreen}\Mem.store(arr + i, v)
  \color{defgreen}\kwelse
    \Map.update(map, i, v)
  \sysprint("put:"+str(i)+str(v)$\!$)
\kwdef get(i:\vint{}) $\equiv$
  \kwvar r

  \color{defgreen}\kwif (0 <= i && i < 100)
    \color{defgreen}r := \Mem.load(arr + i)
  \color{defgreen}\kwelse
    r := \Map.get(map, i)
  \sysprint("get:"+str(i)+str(r)\!)
  \kwreturn r
\end{Verbatim}
\end{minipage}%
\vrule\,%
\begin{minipage}[t]{0.247\textwidth}
$\mwimp{2}$ :=
\begin{Verbatim}[commandchars=\\\{\},codes={\catcode`$=3},fontsize=\small,baselinestretch=0.97]
[\kwmodule \MW{}]
\color{defgreen}\kwlocal first, idx, data
\kwlocal map
\kwdef main() $\equiv$
  \color{defgreen}first := true
  map := \Map.new()
  \App.init()
  \kwwhile (true)
    \App.run()
\kwdef put(i:\vint{},v:\vint{}) $\equiv$
  \color{defgreen}\kwif (first || i == idx) \{
    \color{defgreen}first := false
    \color{defgreen}idx := i
    \color{defgreen}data := v
  \color{defgreen}\} \kwelse
    \Map.update(map, i, v)
  \sysprint("put:"+str(i)+str(v)$\!$)
\kwdef get(i:\vint{}) $\equiv$
  \kwvar r  

  \color{defgreen}\kwif (idx == i)
    \color{defgreen}r := data
  \color{defgreen}\kwelse
    r := \Map.get(map, i)
  \sysprint("get:"+str(i)+str(r)$\!$)
  \kwreturn r
\end{Verbatim}
\end{minipage}%
\vrule\,%
\begin{minipage}[t]{0.253\textwidth}
$\mwmid$ :=
\begin{Verbatim}[commandchars=\\\{\},codes={\catcode`$=3},fontsize=\small,baselinestretch=0.97]
[\kwmodule \MW{}]
\color{defgreen}\kwlocal cls, opt
\kwlocal map
\kwdef main() $\equiv$
  \color{defgreen}cls := opt := (fun _ => 0)
  map := \Map.new()
  \App.init()
  \kwwhile (true)
    \App.run()
\kwdef put(i:\vint{},v:\vint{}) $\equiv$
  \color{defgreen}\kwif (cls(i) == 0)
    \color{defgreen}cls$\,$:=$\,\mapinsert{\code{i}}{\code{choose($\!\{\!1,\!2\!\}\!$)$\!$}}{\code{cls}}$
  \color{defgreen}\kwif (cls(i) == 1)
    \color{defgreen}opt := $\mapinsert{\code{i}}{\code{v}}{\code{opt}}$
  \color{defgreen}\kwelse
    \Map.update(map, i, v)
  \sysprint("put:"+str(i)+str(v)$\!$)
\kwdef get(i:\vint{}) $\equiv$
  \kwvar r
  \color{defgreen}assume(cls(i) != 0)
  \color{defgreen}\kwif (cls(i) == 1)
    \color{defgreen}r := opt(i)
  \color{defgreen}\kwelse
    r := \Map.get(map, i)
  \sysprint("get:"+str(i)+str(r)$\!$)
  \kwreturn r
\end{Verbatim}
\end{minipage}%
\vrule\,%
\begin{minipage}[t]{0.25\textwidth}
$\mwabs$ :=
\begin{Verbatim}[commandchars=\\\{\},codes={\catcode`$=3},fontsize=\small,baselinestretch=0.97]
[\kwmodule \MW{}]
  
\kwlocal full
\kwdef main() $\equiv$

  full := (fun _ => 0)
  \App.init()
  \kwwhile (true)
    \App.run()
\kwdef put(i:\vint{},v:\vint{}) $\equiv$





  full := $\mapinsert{\code{i}}{\code{v}}{\code{full}}$
  \sysprint("put:"+str(i)+str(v)$\!$)
\kwdef get(i:\vint{}) $\equiv$
  \kwvar r




  r := full(i)
  \sysprint("get:"+str(i)+str(r)$\!$)
  \kwreturn r
\end{Verbatim}
\end{minipage}
\hfill\mbox{}
\myhrule
\caption{Two implementations $\mwimp{1}$, $\mwimp{2}$, an intermediate abstraction $\mwmid$, and a full abstraction $\mwabs$ for the module \MW}
\label{fig:mw-def}
\end{figure*}


Contextual refinement (CR) is one of the standard notions of specifying open programs.
For an open program $P$ and a more abstract program $A$ given as its specification,
we say $P$ \emph{contextually refines} $A$, denoted $P \lectx A$, if all possible observable behaviors of $P$
under an arbitrary \emph{closing} context are included in those of $A$ under the same context. 
Here an observable behavior is a terminating or non-terminating trace of observable events
such as input and output events.
An important technical benefit is that CR only requires behaviors of \emph{closed} programs
even though CR relates \emph{open} programs. 

CR has two more advantages: support of $(i)$ horizontal and vertical compositionality
and $(ii)$ full programming features.
First, \emph{horizontal compositionality (HComp)} allows us to
compose modular verification results for different modules, say M1 and M2, as follows:
\[
P_\text{M1} \lectx A_\text{M1} \land P_\text{M2} \lectx A_\text{M2} \implies
P_\text{M1} \link P_\text{M2} \lectx A_\text{M1} \link A_\text{M2}
\]
where $\link$ is the linking operator between modules.
On the other hand, \emph{vertical compositionality (VComp)} allows us to
compose incremental verification results inside the same module, say M, as follows:
\[
P_\text{M} \lectx I_\text{M} \land I_\text{M} \lectx A_\text{M} \implies
P_\text{M} \lectx A_\text{M}
\]
Second, CR imposes no restriction on programming features, so that
it allow cyclic structures such as mutual dependence between modules
and higher-order functions, and also passing pointer values as argument and return values.

However, CR has a downside that it cannot impose conditions on the
context although such conditions are often needed when decomposing a
large contextual refinement into smaller ones.  The reason is because
the definition of CR requires behavioral refinement under \emph{all}
contexts, which indeed plays a crucial role in the general proof of full
compositionality without any restriction on programming features.

In this paper, we address the problem of finding a notion of refinement
that satisfies the three requirements: 
full compositionality, full programming features (in particular, cyclic structures),
and rich conditions on the context.

\paragraph{A motivating example}
\Cref{fig:mw-def} gives a motivating example, 
where the module \MW{} is intended for a simple middleware, and $\mwimp{1}$ and
$\mwimp{2}$ are two (equivalent) implementations for it performing
different optimizations written in \textcolor{defgreen}{green}.

To see what \MW{} does, we look at the common black part of $\mwimp{1}$ and $\mwimp{2}$.
The middleware starts with $\code{main()}$, which creates a
partial map from $\vint$ to $\vint$ by $\code{Map.new()}$, initializes the application by $\code{App.init()}$, and keeps running it by $\code{App.run()}$.
It also provides a map service to the app with \code{put(i,v)},
which maps $\code{i}$ to $\code{v}$ via \code{Map.update} and prints a log,
and \code{get(i)}, which returns the mapped value at the index $\code{i}$ obtained via \code{Map.get} after printing a log.

Then, we see the \textcolor{defgreen}{green parts} for optimizations.
$\mwimp{1}$ optimizes all accesses to the indices between 0 and 100
by storing their data in the array \code{arr} of size 100, allocated by $\code{Mem.alloc(100)}$.
$\mwimp{2}$ optimizes all accesses to the first index given to $\code{put}$
by storing the index and its data in the module-local variables $\code{idx}$ and $\code{data}$. It also uses the variable $\code{first}$ to check whether $\code{put}$ is invoked for the first time.

In order to share the common verification of the black code among $\mwimp{1}$ and $\mwimp{2}$,
we give an intermediate abstraction, $\mwmid$,
that only abstracts the \textcolor{defgreen}{green optimization code} of both $\mwimp{1}$ and $\mwimp{2}$ leaving the black code unchanged.
For this, it uses two mathematical functions $\code{opt}$ and $\code{cls}$ from $\vint$ to $\vint$, where $\code{opt}$ abstracts the optimized storage by mapping the optimized indices to their data and $\code{cls}$ assigns to each index its class:
\code{0} for unused indices, \code{1} for optimized ones, and \code{2} for the rest.
Specifically, \code{put(i,v)} nondeterministically assigns a class to the index \code{i}, if it is unused, via \code{choose(\{1,2\})};
then if \code{i} is an optimized index, updates \code{opt}; otherwise, updates \code{map}.
Then \code{get(i)} assumes%
\footnote{If the assumption fails, it triggers \emph{undefined behavior (UB)} meaning that every possible behavior can happen nondeterministically.}
\code{i} is not unused;
then if it is an optimized index, reads from \code{opt}; otherwise, reads from \code{map}.

Also we give the final abstraction $\mwabs$ that further abstracts the middleware.
It uses one mathematical function $\code{full}$ from $\vint$ to $\vint$, which maps all used indices to their data as implemented in \code{put} and \code{get}.

Then our goal is to \emph{modularly} (\ie separately from other modules) and \emph{incrementally} verify that $\mwimp{1}$ and $\mwimp{2}$ refine $\mwabs$ possibly assuming specific conditions about \emph{external} modules.
Concretely, we wish to modularly verify $(i)$ that $\mwimp{1}$ and $\mwimp{2}$ refine $\mwmid$
by only reasoning about the \textcolor{defgreen}{green optimization code}, and then $(ii)$ that $\mwmid$ refines $\mwabs$.

One of the benefits of this incremental reasoning is that we can factor out and reuse the common verification that $\mwmid$ refines $\mwabs$.
Indeed, if we write another optimization, say $\mwimp{3}$, we just need to verify $\mwimp{3}$ refines $\mwmid$ by only reasoning about the optimization itself.
Moreover, in general, such vertical decomposition can provide nice separation of concerns.
In particular, well decomposed proofs may be more amenable to proof automation~\cite{armada}.

\paragraph{Challenges and existing works}
The key question here is to identify the right notion of modular refinement
(\eg what does it mean for $\mwmid$ to refine $\mwabs$?).
Note that we \mbox{cannot} simply use the (unconditional) contextual refinement \mbox{because} the refinements here are indeed \emph{conditional} on the context:
for example, if \code{Map.get} incorrectly always returns 0, $\mwmid$ cannot refine $\mwabs$ in any sensible way.

Moreover, the intended mutual dependence between \MW{} and \App{} makes it hard to define such conditions since they form (non-monotone) cyclic definitions.
Specifically, since \App{} uses \MW{}, the condition that \App{} behaves well depends on that \MW{} behaves well, which in turn depends
on that \App{} behaves well since \MW{} also uses \App{}.

Although the step-indexing technique~\cite{appel:step-index-original} might be used to solve such cyclic definitions,
it is well known that step-indexed relations are hardly transitive (\ie hardly vertically compositional)~\cite{gil:PB}.
Indeed, among many works using various forms of step-indexed relations~\cite{gil:ml-assembly,reloc,caresl,reloc-reloaded,patterson:funtal,iris-CR,iris-CR2,krebbers:proofmode}, none of them support transitivity of the relations.
Note, however, that some of them~\cite{spies:transfinite,reloc,caresl,reloc-reloaded,iris-CR,iris-CR2} use step-indexed relations as a means to establish the unconditional CR, which allows VComp but no conditions on contexts.

Another line of work such as CAL (Certified Abstraction Layers)~\cite{gu:dscal} and refinement calculus~\cite{back2012refinement}
avoids such cyclic definitions by disallowing those programming features that may introduce cyclicity
such as mutual dependence and pointer-value passing between different modules.
Indeed, CertiKOS~\cite{CertiKOS16}, an operating system verified with CAL, clearly shows
the merit and limitation of CAL.
CertiKOS enjoys HComp and VComp of CAL by modularly and \mbox{incrementally} decomposing the whole refinement into 74 sub-refinements~\cite{gu:ccal, gu:ccal-dev}.
However, it could not use any dynamic memory allocation since it would require pointer-value passing, and instead had to only use static variables of fixed sizes allocated at the booting time.
Koenig and Shao~\cite{jeremie:lics,jeremie:thesis} presents a generalization of CAL that supports mutual recursion; however it still lacks support of rich conditions expressing various ownership as in modern separation logics~\cite{iris2015,VST}.

Yet another line of work such as parametric bisimulation~\cite{gil:PB}, RGSim (Rely-Guarantee-based Simulation)~\cite{feng:rgsim} and RUSC (Refinement Under Self-related Contexts)~\cite{CompCertM} supports only a limited form of conditions on context,
although fully supporting compositionality and programming features.
More precisely, they do not allow specific conditions per module but only global conditions that equally apply to every module.

To sum up, to our best knowledge, no existing work can verify our motivating example \emph{modularly and incrementally} as outlined above.

\paragraph{Our approach}
As a solution, we propose CCR~(Conditional Contextual Refinement),
which uses the (unconditional) contextual refinement as an underlying notion
but overcomes its shortcoming by encoding module-specific conditions as executable code added to the abstract module.
We will outline how CCR works for the motivating example.

First, although $\mwimp{1} \lectx \mwmid$ does not hold,
we can make $\mwimp{1} \lectx {} \mwmidcond$ hold for:
\[\small
\begin{array}{@{}l@{}}
  \mwmidcond ~\defeq~ \mwmid + \Grt{\Mem{} is used well} + \Asm{\Mem{} behaves well} \\
  \quad {} + \Asm{\MW{} is used well, simply} + \Grt{\MW{} behaves well, simply}
\end{array}  
\]
Specifically, $\mwmidcond$ is $\mwmid$ extended with four kinds of executable code
capturing the needed conditions.
The two \GrtL{}'s encode guarantees about \MW{}'s behavior saying that \MW{} uses \Mem{} well and also behaves well, simply;
and the two \AsmL{}'s encode assumptions about \Mem{}'s behavior saying that \Mem{} uses \MW{} well, simply,
and also behaves well.

There are a few points to note.
First, we do not need any conditions about \Map{} and \App{} since the black code using them is not abstracted in this refinement.
Second, we only need a simple condition about \MW{} for interaction with \Mem{}, which is why we put ``simply''.
The condition basically says that \Mem{} does not use \MW{}.
A full condition, including the guarantee that \code{MW.put} and \code{MW.get} behave like a map assuming they are used well,
will be needed to reason about interaction with \App{} in the next abstraction.

The most important point here is that technically we assume nothing about external modules and thus do not need any cyclic definitions.
Indeed, \mwimp{1} refines $\mwmidcond$ even when linked with badly-behaved \Mem{}, which is why the contextual refinement holds.
The intuition is that \AsmL{} (or \GrtL{}) checks whether what has been observed by \MW{} is consistent with the \asm{assumed} (or \grt{guaranteed}) behaviors;
then if the \asm{assumption} (or \grt{guarantee}) fails, it triggers \asm{UB} (or \grt{NB}).
Here UB is called \emph{undefined behavior} and interpreted as triggering every possible behavior nondeterministically, so that the refinement holds trivially if UB is triggered;
that is, we just need to establish the refinement only when UB does not occur.
On the other hand, NB is called \emph{no behavior} and, roughly speaking, interpreted as empty behavior, so that
the refinement cannot hold if NB is triggered;
that is, in order to establish the refinement, we have to prove that NB does not occur.

Also we can prove $\mwimp{2} \lectx \mwmidcond$ and $\mwmidcond \lectx \mwabscond$ for:
\[\small
\begin{array}{@{}l@{}l@{}}
  \mwabscond ~\defeq~ \mwabs
  & {} + \Grt{\Mem{} is used well} + \Asm{\Mem{} behaves well} \\
  & {} + \Grt{\Map{} is used well} + \Asm{\Map{} behaves well} \\
  & {} + \Grt{\App{} is used well} + \Asm{\App{} behaves well} \\
  & {} + \Asm{\MW{} is used well} + \Grt{\MW{} behaves well} \\
\end{array}  
\]
By VComp of $\lectx$, we have $\mwimp{i} \lectx \mwabscond$ for $i\in \setofz{1,2}$.

For other modules, we can prove the following possibly incrementally:
\[
\memimp \lectx \memabscond, \qquad
\mapimp \lectx \mapabscond, \qquad
\appimp \lectx \appabscond
\]
where
\[
\small
\begin{array}{@{}l@{}l@{}}
  \memabscond ~\defeq~ \memabs
  & {} + \Asm{\Mem{} is used well} + \Grt{\Mem{} behaves well} \\
  \mapabscond ~\defeq~ \mapabs
  & {} + \Grt{\Mem{} is used well} + \Asm{\Mem{} behaves well} \\
  & {} + \Asm{\Map{} is used well} + \Grt{\Map{} behaves well} \\
  \appabscond ~\defeq~ \appabs
  & {} + \Grt{\MW{} is used well} + \Asm{\MW{} behaves well} \\
  & {} + \Asm{\App{} is used well} + \Grt{\App{} behaves well} \\
\end{array}  
\]
Note that we have the conditions as above since \memimp uses no external module,
\mapimp uses only \Mem{}, and \appimp uses only \MW{}.
Then, by HComp of $\lectx$, we have: for $i\in \setofz{1,2}$,
\[
 \mwimp{i} \link \memimp \link \mapimp \link \appimp \lectx \mwabscond \link \memabscond \link \mapabscond \link \appabscond
\]

Finally, we have the following refinement thanks to our \emph{Assumption Cancellation Theorem (ACT)} for a closed program,
which cancels out each \AsmL{} in the matching \GrtL{} and then freely eliminates all \GrtL{}'s by definition.
\[
\mwabscond \link \memabscond \link \mapabscond \link \appabscond \lectx \mwabs \link \memabs \link \mapabs \link \appabs
\]
Again, by VComp of $\lectx$, we have: for $i\in \setofz{1,2}$,
\[
 \mwimp{i} \link \memimp \link \mapimp \link \appimp \lectx \mwabs \link \memabs \link \mapabs \link \appabs
\]
Note that conditions in intermediate abstractions such as \Grt{\MW{} behaves well, simply}
are never used by the cancellation theorem,
so that they can be freely chosen regardless of the actual conditions used by external modules.

To summarize, we addressed the challenging problem by developing a novel mechanism (\ie \AsmL{} and \GrtL{}) to \emph{operationally} and \emph{module-locally} encode rich conditions on context modules.
Moreover, \AsmL{} and \GrtL{} are \emph{auto-generated} from propositions that can express various ownership via PCMs (Partial Commutative Monoids)~\cite{pym:PCM-original,calcagno2007local} as in the state-of-the-art separation logics (SLs) such as \iris~\cite{iris2015} and \vst~\cite{VST}.
Unlike those SLs, we encode PCM-based ownership without using step-indexing even for mutually dependent modules, which is essential to support VComp.
Also, we believe that the key ideas of CCR could be applied to the verification of CertiKOS to lift its current restrictions (\ie the absence of dynamic allocation and mutual dependence).

\paragraph{Contributions}
All our results are fully formalized in Coq~\cite{supplementary} and summarized as follows.

\textbf{(1)} We develop the first theory, CCR, that provides a notion of refinement supporting full compositionality, full (sequential) programming features and rich conditions.

\textbf{(2)} We develop \ems~(Executable Module Semantics) as a general underlying module semantics for CCR,
which is uni-typed with the most general type consisting of all mathematical values,
and allows each module to be equipped with an arbitrary small-step operational semantics (expressed in terms of interaction trees~\cite{liyao:itree})
with a given set of events such as primitive events and calls to external functions.
\ems also supports two notions of nondeterminism, called \emph{demonic} and \emph{angelic} one in the literature~\cite{bodik2010programming, back2012refinement, tyrrell2006lattice, jeremie:lics}, which we crucially use to encode rich conditions on context modules.

\ems has two advantages.
First, many languages with different type systems can be embedded into \ems since, in particular, typing assumptions or guarantees about argument and return values can be easily expressed.
As examples, we develop the following languages and embed them into \ems.
\begin{itemize}[leftmargin=8pt]
\item \imp: A C-like language with integer and (function and memory) pointer values, which is used to write implementation code and compiled down to assembly via our verified compiler for \imp together with CompCert.
\item \originaltxt{\spc: A specification language, which is used to write conditions such as \AsmL{}'s and \GrtL{}'s above.}
  \spc: A specification language in which one can specify various conditions such as \AsmL{}'s and \GrtL{}'s above and also write abstract yet executable code.
\end{itemize}
Second, since modules in \ems are written as interaction trees, they are executable via Coq's extraction mechanism into OCaml, so that we can test them.



\textbf{(3)} We verify various examples (including the motivating example) written in \imp,
which demonstrates
$(i)$ modular and incremental verification via intermediate abstractions (\ie full compositionality);
$(ii)$ cyclic and higher-order reasoning about recursion and function pointers (\ie full features);
and $(iii)$ reasoning about PCM-based ownership (\ie rich conditions).

For this verification,
we use the CCR proof mode in Coq supporting both $(i)$ simulation reasoning about preservation of side effects such as IO events and termination and $(ii)$ propositional reasoning about conditions on contexts, for which we employ the IPM (\iris Proof Mode) package~\cite{krebbers:proofmode} (\ie by instantiating it with our CCR theory) that streamlines reasoning about PCMs.
Note that the details of the simulation technique and CCR proof mode
will be published elsewhere.


\textbf{(4)} We develop a \emph{verified} compiler for \imp targeting Csharpminor of CompCert~\cite{CompCert}, which is in turn compiled by the \mbox{verified} compiler CompCert to generate assembly code.
As a result, we formally establish behavioral refinement from the top-level abstractions of the above examples, all the way down to their compiled assembly code.


\section{Key ideas}
\label{sec:overview}

\subsection{Technical challenges and our solution}

To understand the challenges with defining \AsmL{} and \GrtL{},
we give an example, where we want to abstract $\onceimp$ into $\onceabs$.
\[
\begin{minipage}{0.45\columnwidth}
\begin{Verbatim}[commandchars=\\\{\},codes={\catcode`$=3},fontsize=\small,baselinestretch=0.95]
\onceimp:=[\kwmodule \Once{}]
\kwlocal done := false
\kwdef do() $\equiv$
  \kwif (done) \sysprint("err")
  \kwelse done := true
\end{Verbatim}
\end{minipage}\vrule\ \ %
\begin{minipage}{0.42\columnwidth}
\begin{Verbatim}[commandchars=\\\{\},codes={\catcode`$=3},fontsize=\small,baselinestretch=0.95]
\onceabs:=[\kwmodule \Once{}]

\kwdef do() $\equiv$
  \kwskip

\end{Verbatim}
\end{minipage}
\]
The question here is how to encode and add the condition \Asm{\code{do()} hasn't been called}
to $\code{do()}$ in $\onceabs$,
in a way that if a client locally proves \Grt{\code{do()} hasn't been called} before calling \code{do()},
then they are canceled out when linked.

A naive encoding $\onceabstry$ might be as follows.
\[
\begin{minipage}{0.44\columnwidth}
\begin{Verbatim}[commandchars=\\\{\},codes={\catcode`$=3},fontsize=\small,baselinestretch=0.95]
\onceabstry:=[\kwmodule \Once{}]
\asm{\kwlocal done := false}
\kwdef do() $\equiv$
  \asm{\kwassume(done $=$ false)}
  \asm{done := true}
\end{Verbatim}
\end{minipage}\vrule\ \ \ %
\begin{minipage}{0.54\columnwidth}
\begin{Verbatim}[commandchars=\\\{\},codes={\catcode`$=3},fontsize=\small,baselinestretch=0.95]
\testimp{1}{}$=$\testabs{1}:=[\kwmodule{} \Test{1}]
\kwdef main() $\equiv$ \Once{}.do()
\end{Verbatim}
\hrule{}
\begin{Verbatim}[commandchars=\\\{\},codes={\catcode`$=3},fontsize=\small,baselinestretch=0.95]
\testabscond{1}:=[\kwmodule \Test{1}]
\kwdef main() $\equiv$
  \Grt{...}; \Once{}.do()
\end{Verbatim}
\end{minipage}
\]
We can easily see that $\onceimp \lectx \onceabstry$ holds since, when called twice, 
\code{do()} will trigger \asm{UB} by \code{\asm{\kwassume(done $=$ false)}} rendering all possible behaviors.

However, it is unclear how to state the guarantee condition \code{\Grt{...}} in $\testabscond{1}$ above,
so that it can automatically cancel out \code{\asm{\kwassume(done $=$ false)}} in $\onceabstry$ when linked.
The problem is that the module \Test{1} cannot access the local variable \code{done} of \Once{}
and even worse, since the context module \Once{} is arbitrary in $\testimp{1} \lectx \testabscond{1}$,
it may not have such a local variable at all.

To understand the requirement for the encoding more clearly,
consider the following variation.
\[
\begin{minipage}{0.56\columnwidth}
\begin{Verbatim}[commandchars=\\\{\},codes={\catcode`$=3},fontsize=\small,baselinestretch=0.95]
\testimp{2}{}$=$\testabs{2}:=[\kwmodule \Test{2}]
\kwdef main() $\equiv$
  \Once{}.do()
  \Once{}.do()
\end{Verbatim}
\end{minipage}\vrule\ \ %
\begin{minipage}{0.42\columnwidth}
\begin{Verbatim}[commandchars=\\\{\},codes={\catcode`$=3},fontsize=\small,baselinestretch=0.95]
\testabscond{2}:=[\kwmodule \Test{2}]
\kwdef main() $\equiv$
  \Grt{...}; \Once{}.do()
  \Grt{...}; \Once{}.do()
\end{Verbatim}
\end{minipage}
\]
It is clear that something should go wrong here because \code{\Once{}.do()} is called twice.
Here what we desire is that
since actual proofs should be done module-locally in $\onceimp \lectx \onceabstry$ and $\testimp{2} \lectx \testabscond{2}$,
the cancellation of syntactically matched \AsmL{} and \GrtL{} should hold unconditionally without requiring reasoning about the conditions;
that is, the following should hold:
\[
\onceabstry \link \testabscond{2} \lectx \onceabs \link \testabs{2}
\]
Instead, the local reasoning of $\testimp{2} \lectx \testabscond{2}$ should go wrong since
\code{\Test{2}.main()} is blamable due to its two calls of \code{\Once{}.do()}.
From this, it follows that the second \code{\Grt{...}} in $\testabscond{2}$ should effectively prevent the second call to \code{\Once{}.do()}.
The reason is because invoking \code{\Once{}.do()} twice triggers \asm{UB} in $\onceabstry \link \testabscond{2}$ while $\onceabs \link \testabs{2}$ never triggers \asm{UB}.

To sum up, the challenge here is that
the \AsmL{} in \code{\Once{}.do()} and the \GrtL{} in \code{\Test{i}.test()} are completely independent and local computations
since there is no argument passing or secrete channel;
however, they should affect each other and prevent undesired computation in the right place.

\begin{figure}[t]
\begin{minipage}[t]{0.43\columnwidth}
\begin{Verbatim}[commandchars=\\\{\},codes={\catcode`$=3},fontsize=\small,baselinestretch=0.95]
$\onceabscond$:=[\kwmodule \Once{}]
\boilerplate{\kwlocal \mres := $\munit$}\phantom{$\stackrel{\text{=}}{=}$}
\kwdef do() $\equiv$
  \asm{\kwvar frm :=}
    \Asm{$\lambda$\res.$\;$\res${}\ge{}$\Do, \kwemp}
  \kwskip
  \Grt{$\lambda$\res.$\;$True, frm}
\end{Verbatim}
\end{minipage}%
\vrule\,%
\begin{minipage}[t]{0.5\columnwidth}
\begin{Verbatim}[commandchars=\\\{\},codes={\catcode`$=3},fontsize=\small,baselinestretch=0.95]
$\testabscond{i}$:=[\kwmodule \Test{i}]
\boilerplate{\kwlocal \mres := $\munit$}\phantom{$\stackrel{\text{=}}{=}$}
\kwdef main() $\equiv$
  \asm{\kwvar frm :=} \Asm{$\lambda$\res.$\;$\res${}\ge{}$\Do, \kwemp}
  \kwrepeat $i$ \{
    \grt{frm :=} \Grt{$\lambda$\res.$\;$\res${}\ge{}$\Do, frm}
    \Once{}.do()
    \asm{frm :=} \Asm{$\lambda$\res.$\;$True, frm} \}
  \Grt{$\lambda$\res.$\;$True, frm}
\end{Verbatim}
\end{minipage}%
\hfill\mbox{}\\[2mm]%
\hrule\mbox{}\\
\begin{minipage}[t]{0.48\columnwidth}
\begin{Verbatim}[commandchars=\\\{\},codes={\catcode`$=3},fontsize=\small,baselinestretch=0.95]
\color{purple}\Asm{Cond, \lres} $\equiv$ \{

\color{purple} \kwvar$\,$(\eres,\res)$\,$:=$\,$\kwtake($\Sigma\times\Sigma$)
\color{purple} \kwassume(Cond \res)
\color{purple} \kwassume$(\mval(\mres+\lres+{}$
\color{purple}          $\eres+\res))$
\color{purple} \eres \}
\end{Verbatim}
\end{minipage}%
$\,\,$%
\begin{minipage}[t]{0.48\columnwidth}
\begin{Verbatim}[commandchars=\\\{\},codes={\catcode`$=3},fontsize=\small,baselinestretch=0.95]
\color{blue}\Grt{Cond, \eres} $\equiv$ \{
\color{blue} \mres := \kwchoose($\Sigma$)
\color{blue} \kwvar$\,$(\lres,\res)$\,$:=$\,$\kwchoose($\Sigma\times\Sigma$)
\color{blue} \kwguarantee(Cond \res)
\color{blue} \kwguarantee$(\mval(\mres+\lres+{}$
\color{blue}             $\eres+\res))$
\color{blue} \lres \}
\end{Verbatim}
\end{minipage}%
\hfill\mbox{}
\myhrule
\caption{Conditional abstractions $\onceabscond$, $\testabscond{1}$, $\testabscond{2}$}
\label{fig:once-def}
\end{figure}


\paragraph{Our solution}
\Cref{fig:once-def} shows how we encode the conditions for $\onceabs$ and $\testabs{i}$ for $i\in\setofz{1,2}$.
Before we proceed, we remark an interesting point.
In fact, the two conditional abstractions $\onceabstry$ and $\onceabscond$ are \emph{contextually equivalent}:
\[
\onceabstry \lectx \onceabscond  \quad\text{ and }\quad \onceabscond \lectx \onceabstry
\]
Thus the following contextual refinements are equivalent.
\[
\begin{array}{@{}l@{}l@{}}
&\onceabscond \link \testabscond{i} \lectx \onceabs \link \testabs{i}\\[1mm]
{}\iff{}&\onceabstry \link \testabscond{i} \lectx \onceabs \link \testabs{i}
\end{array}
\]
However, the former is much easier to prove than the latter.
Specifically,
the former proof just requires to check that the \AsmL{}'s and \GrtL{}'s are syntactically matched
without requiring any reasoning about the conditions specified in them,
while the latter proof essentially requires non-trivial reasoning about the conditions. 
It is important to note that such difference is made due to the specific ways they are computed, not due to their observable behaviors.

Inspired by the approach of modern SLs~\cite{VST, iris2015},
we also encode various ownership representing certain \emph{capability} and \emph{knowledge}
using partial commutative monoids (PCMs),
but \emph{operationally} and \emph{module-locally} (\ie independently from external modules).
Note that a PCM $\Sigma$ is a set equipped with a commutative and associative binary operator $+$ on $\Sigma$, called \emph{addition}, an identity element $\munit$,
and a validity predicate $\mval$ on $\Sigma$ satisfying $(i)$ $\mval(\munit)$ and $(ii)$ $\forall a, b.~ \mval(a\madd b) \implies \mval(a)$.
Since invalid elements are considered \emph{undefined}, $+$ can be seen as a partial operator.
Elements of $\Sigma$ are called \emph{resources} and we say $a \ge b$ for $a,b\in\Sigma$ if $\exists c\in\Sigma.~ a = b + c$.

In $\onceabscond$, we add conditions basically saying that \code{do()} should be invoked at most once.
For this, we can use any PCM $\Sigma$ involving a resource, \Do{}, such that
$\Do + \Do$ is \emph{undefined} (\ie whose result is invalid), which represents inconsistency.
Then, $\Do$ represents the capability to invoke \code{do()}
and we can see that there is at most one \Do{} since $\Do+\Do$ is undefined.

With this intuition, we look at \code{do()} in $\onceabscond$.
The \AsmL{} at the beginning encodes the assumption that $\Do$ is given to \code{do()}
by the caller, which forms \Asm{\code{do()} hasn't been called};
the \GrtL{} at the end says that we guarantee nothing.
The special module-local variable $\mres$ is only used by \AsmL{} and \GrtL{} and initialized with a specific resource for each module, which is $\munit$ for every module in this example.
The intuition here is that invoking \code{do()} requires the resource \Do{},
which is consumed (\ie not returned) so that \code{do()} cannot be invoked anymore.
Similarly, \code{main()} requires \Do{} initially and
\emph{guarantees} to give \Do{} before each invocation of \code{\Once.do()}.

Now we see the computational interpretation of \AsmL{} and \GrtL{},
whose definitions are given at the bottom of \Cref{fig:once-def}.
Note that each function of a module (decorated with conditions) maintains four kinds of resources
at every interaction point with a caller/callee:
\begin{itemize}[leftmargin=8pt]
\item a \emph{module resource} stored in the module-local variable $\mres$ that is owned by the module (\ie shared among all functions in the module),
\item a \emph{local resource}, $\lres$, that is locally owned by the current invocation of the function,
\item an \emph{external resource}, $\eres$, that is conceptually the summation of all the other local and module resources in the whole system at the moment,
\item a \emph{call resource}, $\res$, that is conceptually passed from the caller/callee or will be passed to the caller/callee.
\end{itemize}

Then, \code{\Asm{Cond, $\lres$}} for \code{Cond} a predicate on
$\Sigma$ and $\lres$ the current local resource, which is $\munit$ at
the beginning of each function invocation, computes as follows.
It magically \emph{takes} an external resource $\eres$ and a call resource~$\res$ via $\kwtake$
(which are conceptually passed from somewhere but technically out-of-thin-air, which we will explain soon);
\emph{assumes} \code{(Cond~\res)} holds (\ie if unsuccessful, triggers \asm{UB} rendering all possible behaviors);
\emph{assumes} all four resources are consistent (\ie their summation is valid);
and returns $\eres$.
As in \code{do()} and \code{main()},
the returned $\eres$ is stored at the variable $\frm$ and passed to the next \GrtL{}.

Then, \code{\Grt{Cond, $\eres$}} for $\eres$ the current external resource
computes as follows.
It nondeterministically \emph{chooses} a module resource and updates the module-local variable $\mres$ with it
and further \emph{chooses} a local resource $\lres$ and a call resource $\res$ via $\kwchoose$ (which are conceptually passed to somewhere but technically nowhere, which we will explain soon);
\emph{guarantees} \code{(Cond \res)} holds (\ie if unsuccessful, triggers \grt{NB}, called \emph{no behavior}, basically rendering empty behavior);
\emph{guarantees} all four resources are consistent;
and returns $\lres$.
The returned $\lres$ is stored at the variable $\frm$ and passed to the next \AsmL{}.

Now we see how \kwtake{} and \kwchoose{} can make an illusion of receiving and sending any ghost information, including a resource, from and to a caller/callee.
Indeed, a function cannot physically receive or send any ghost information in the setting of contextual refinement because
context modules are completely arbitrary so that they may well be just physical modules such as those written in \imp.
However, by using the standard (a.k.a. demonic) nondeterministic choice function \kwchoose{} and its dual (a.k.a. angelic~\cite{bodik2010programming, back2012refinement, tyrrell2006lattice, jeremie:lics}) one \kwtake{}, we can logically make such an illusion.
Basically, we define the operational behaviors of \kwchoose{}($X$) and \kwtake{}($X$)
for any set $X$ as follows:
\[
\begin{array}{@{}l@{\ \ }c@{\ \ }l@{}}
  \beh{\code{\kwvar{} x := \kwchoose($X$); $K[\code{x}]$}} &\defeq& \bigcup_{x\in X} \beh{K[x]} \\
  \beh{\code{\kwvar{} x := \kwtake($X$); $K[\code{x}]$}} &\defeq& \bigcap_{x\in X} \beh{K[x]} \\
\end{array}
\]
%
It is important to note that 
even though defining angelic nondeterminism for \emph{open} programs is non-trivial~\cite{jeremie:lics},
we can avoid such a problem and give a simple operational semantics as above since we only need to define it for \emph{closed} programs, thanks to the use of contextual refinement.

\paragraph{The example revisited}
Then we discuss how such an illusion can actually make refinement between $\onceimp$ and $\onceabscond$ hold for \emph{any} context including badly behaved ones (\ie $\onceimp \lectx \onceabscond$).
When \code{do()} is invoked for the first time,
we need to establish the refinement for \emph{any} taken $\eres$ and $\res$ that satisfy both
$\kwassume$ in \Asm{$\lambda$\res.$\,$\res${}\ge{}$\Do{},$\,$\kwemp} by definition of \kwtake{}.
Note that an unsuccessful $\kwassume$ triggers \asm{UB} rendering all possible behaviors and thus the refinement trivially holds.
Now, for any such taken $\eres$ and $\res$, both $\onceimp$ and $\onceabscond$ return,
where we successfully \emph{choose} to update $\mres$ with $\Do$ by executing \Grt{\code{$\lambda$\res.$\,$True,$\,$frm}} since the \code{frm} given by \Asm{$\lambda$\res.$\,$\res${}\ge{}$\Do{},$\,$\kwemp} cannot include $\Do$.
Since then, whenever \code{do()} is invoked again (badly),
the refinement trivially holds since $\mres$ contains $\Do$ and thus
\Asm{$\lambda$\res.$\,$\res${}\ge{}$\Do{},$\,$\kwemp} must trigger \asm{UB} in $\onceabscond$
since the condition cannot be met.
Note that this informal argument can be made formal by using a simple simulation relation
between $\onceimp$ and $\onceabscond$ with the following relational invariant:
\[
  (\code{done}=\code{false} \land \mres = \munit)
  \lor{} (\code{done}=\code{true} \land \mres = \Do)
\]

Similarly, one can easily see $\testimp{1} \lectx \testabscond{1}$ holds.

Now we discuss how $\testabscond{2}$ prevents the second call to \code{\Once{}.do()}.
By the definition of \kwtake{}, it suffices to show the prevention (\ie triggering NB before the call) for some \frm{} that is \emph{successfully taken} by \Asm{$\lambda$\res.$\,$\res${}\ge{}$\Do{},$\,$\kwemp}.
Here we take $(\eres,\res)$ to be $(\munit,\Do)$, which satisfies both \kwassume{} in the \AsmL{}.
Then, we have to show the prevention of the second call for any successful resources \emph{chosen} from
\Grt{$\lambda$\res.$\,$\res${}\ge{}$\Do{},$\,$\frm} by the definitions of \kwchoose{} and \kwguarantee{}.
Here we can see that the chosen \frm{} cannot contain \Do{} since $\res$ contains it.
Then after the first call,
we successfully take \frm{} to be \Do{} by \Asm{$\lambda$\res.$\,$True,$\,$\frm} since the previous \frm{} cannot contain \Do{}.
Then before the second call,
\Grt{$\lambda$\res.$\,$\res${}\ge{}$\Do{},$\,$\frm} must trigger \grt{NB}
since \frm{} contains \Do{} and thus the condition cannot be met.
As a result, since \grt{NB} renders empty behavior, the second call cannot be executed.

From this prevention, $\testimp{2} \not\lectx \testabscond{2}$ also follows.

\vspace*{-0.6mm}
\paragraph{Assumption Cancellation}
Finally, we discuss how the assumptions are canceled out by the matching guarantees
when the two conditional abstractions are linked:
\[
\onceabscond \link \testabscond{i} \lectx \onceabs \link \testabs{i}
\]
The assumption cancellation theorem (ACT) is proven in general for any \emph{closed} program as follows.

To \emph{discharge} the initial \AsmL{} of $\main()$ (\ie replacing it with \kwskip{}),
we just need to take an initial resource $\res$ to $\main()$, here $\Do$, 
that $(i)$ is consistent with 
all the initial module resources,
here $\munit$ for both $\onceabscond$ and $\testabscond{i}$;
and $(ii)$ satisfies the precondition of $\main()$, here \mbox{$\res\ge\Do$}.
For $\eres$, we take the summation of the initial resources of all external modules.

Then, every subsequent \AsmL{} is automatically discharged by carefully taking its resources depending on the resources chosen by the matching (\ie immediately preceding) \GrtL{}.
Specifically, we take $(i)$ $\res$ to be the same $\res$ chosen by the \GrtL{} and $(ii)$ $\eres$ to be the summation of all the local and module resources in the system except for its own $\lres$ and $\mresb$.
Then the two \kwassume{}'s in the \AsmL{} are discharged by the two \kwguarantee{}'s in the \GrtL{}
because $\mresb + \lres + \eres$ in \AsmL{} is equal to that in \GrtL{}.
This is because both summations coincide with the summation of all the local and module resources in the system by construction.


We conclude with two remarks.
First, in our Coq formalization, we slightly generalize the CCR framework to support the ACT theorem for \emph{open} programs,
which cancels out every assumption by the matching guarantee even in the presence of unverified (logically orthogonal) contexts.
Second, even after applying the ACT theorem (in particular to a selected set of verified modules using the generalized CCR),
we can further abstract the resulting abstraction (possibly together with other unverified modules) using the (generalized) CCR.
The 
technical report~\cite{supplementary} presents the details of the generalized CCR framework and
the examples for the above two cases in \S3.1 and \S3.3, respectively.


\vspace*{-0.6mm}
\subsection{Incremental verification}

\begin{figure*}[t]
\begin{minipage}[t]{0.24\textwidth}
$\appimp$ :=
\begin{Verbatim}[commandchars=\\\{\},codes={\catcode`$=3},fontsize=\small,baselinestretch=0.95]
[\kwmodule \App{}]
\color{defgreen}\kwlocal initialized := false
\kwdef init() $\equiv$
  \color{defgreen}\kwif (initialized) \{
    \color{defgreen}\sysprint("error: init")
  \color{defgreen}\} \kwelse \{
    \color{defgreen}initialized := true
    \MW.put(0, 42)
  \color{defgreen}\}

\kwdef run() $\equiv$
  \color{defgreen}\kwif (!initialized) \{
    \color{defgreen}\sysprint("error: run")
  \color{defgreen}\} \kwelse \{

    \kwvar v := \MW.get(0)

    \sysprint("val:"+str(v))
  \color{defgreen}\}
\end{Verbatim}
\end{minipage}%
\vrule\,%
\begin{minipage}[t]{0.24\textwidth}
$\appmidcond$ :=
\begin{Verbatim}[commandchars=\\\{\},codes={\catcode`$=3},fontsize=\small,baselinestretch=0.95]
[\kwmodule \App{}]
\boilerplate{\kwlocal \mres := \Run}
\kwdef init() $\equiv$
  \asm{\kwvar frm :=}
    \Asm{$\lambda$\res. \res $\ge$ \Init, \kwemp}

  \grt{frm :=} \Grt{$\lambda$\res.$\,$True, frm}
  \MW.put(0, 42)
  \asm{frm :=} \Asm{$\lambda$\res.$\,$True, frm}
  \Grt{$\lambda$\res. \res $\ge$ \Run, frm}
\kwdef run() $\equiv$
  \asm{\kwvar frm :=}
    \Asm{$\lambda$\res. \res $\ge$ \Run, \kwemp}

  \grt{frm :=} \Grt{$\lambda$\res.$\,$True, frm}
  \kwvar v := \MW.get(0)
  \asm{frm :=} \Asm{$\lambda$\res.$\,$True, frm}
  \sysprint("val:"+str(v))
  \Grt{$\lambda$\res. \res $\ge$ \Run, frm}
\end{Verbatim}
\end{minipage}%
\vrule\,%
\begin{minipage}[t]{0.25\textwidth}
$\appabscond$ :=
\begin{Verbatim}[commandchars=\\\{\},codes={\catcode`$=3},fontsize=\small,baselinestretch=0.95]
[\kwmodule \App{}]
\boilerplate{\kwlocal \mres := \Run}
\kwdef init() $\equiv$
  \asm{\kwvar fi := \kwtake($\vint\,\rightarrow\,\vintopt$)}
  \asm{\kwvar frm := \Asm{$\lambda$\res. \res $\ge$ \Init{}$\;$+$\;$\MWis{}(fi), \kwemp}}
  \grt{\kwvar fe := \kwchoose($\vint\,\rightarrow\,\vintopt$)}
  \grt{frm :=} \Grt{$\lambda$\res. \res $\ge$ \MWis{}(fe), frm}
  \MW.put(0, 42)
  \asm{frm :=} \Asm{$\lambda$\res. \res $\ge$ \MWis{}($\mapinsert{\code{0}}{\code{42}}{\code{fe}}$), frm}
  \Grt{$\lambda$\res. \res $\ge$ \Run{}$\;$+$\;$\MWis{}($\mapinsert{\code{0}}{\code{42}}{\code{fi}}$), frm}
\kwdef run() $\equiv$
  \asm{\kwvar fi := \kwtake($\vint\,\rightarrow\,\vintopt$)}
  \asm{\kwvar frm$\;$:=$\;$\Asm{$\lambda$\res.$\;$\res$\;\ge\;$\Run{}$\;$+$\;$\MWis{}(fi)$\;\land\;$fi(0)${}={}$\some{}$\,$42, \kwemp}}
  \grt{\kwvar (fe, ve) := \kwchoose(($\vint\,\rightarrow\,\vintopt$) $\times \vint$)}
  \grt{frm :=} \Grt{$\lambda$\res. \res $\ge$ \MWis{}(fe) $\land$ fe(0)${}={}$\some{}$\;$ve, frm}
  \kwvar v := \MW.get(0)
  \asm{frm :=} \Asm{$\lambda$\res. \res $\ge$ \MWis{}(fe) $\land$ v${}={}$ve, frm}
  \sysprint("val:"+str(42))
  \Grt{$\lambda$\res. \res $\ge$ \Run{}$\;$+$\;$\MWis{}(fi) $\land$ fi(0)${}={}$\some{}$\,$42, frm}
\end{Verbatim}
\end{minipage}%
\hfill\mbox{}
\myhrule
\caption{An implementation $\appimp$, an intermediate conditional abstraction $\appmidcond$, a full conditional abstraction $\appabscond$ for \App}
\label{fig:app-def}
\end{figure*}


To present the key ideas behind incremental verification,
we give a simple implementation $\appimp$, shown in \Cref{fig:app-def},
for the module \App{} of the motivating example.
We incrementally verify it against $\appabs$ via $\appmid$ just for presentation purposes.

In $\appimp$, the \code{init()} function first checks whether \App{} is already initialized or not using the module-local variable \code{initialized};
if so, prints an error; otherwise, assigns \code{true} to the variable and put \code{42} at index \code{0} using \code{\MW.put}.
The \code{run()} function also checks whether \App{} is initialized;
if not, prints an error; otherwise, fetches the value at index \code{0} using \code{\MW.get} and prints it.
In $\appmidcond$, we abstract away the \textcolor{defgreen}{green error checking code} of $\appimp$, and
add the following conditions:
\[\small
\Asm{\App{} is used well, simply} + \Grt{\App{} behaves well, simply}
\]
Note that we do not need any condition on \MW{} since the black code using \MW{} is not abstracted.
In $\appabscond$, we further optimize the function \code{run} by replacing the fetched value \code{v} with \code{42}
and add the following conditions:
\[\small
\begin{array}{l}
\Grt{\MW{} is used well} + \Asm{\MW{} behaves well}+{} \\
\Asm{\App{} is used well} + \Grt{\App{} behaves well}
\end{array}
\]

\paragraph{First abstraction}
In $\appmidcond$, we add conditions basically saying that \code{init()} should be invoked once and then \code{run()} can be invoked many times.
For this, we can use any PCM $\Sigma$ involving three resources, \Init{}, \Run{} and \Both{}, with the following law:
\[
\Init + \Run = \Both
\]
All the other additions among them are \emph{undefined} (\ie whose results are invalid), which represents inconsistency.
Then, $\Init$ and $\Run$ represent the capability to invoke \code{init()} and \code{run()}, respectively.
From the laws of $+$, we can see that there are at most one \Init{} and one \Run{} since
$\Init+\Init$ and $\Run+\Run$ are undefined.

With this intuition, we look at \code{init()} and \code{run()} in $\appmidcond$.
The \AsmL{}'s at the beginning of them encode the assumptions that $\Init$ / $\Run$ is given to \code{init()} / \code{run()} respectively by the caller,
which form \Asm{\App is used well, simply}.
On the other hand,
the \GrtL{}'s at the end of them encode the guarantees that $\Run$ is returned to the caller of \code{init()} / \code{run()},
which form \Grt{\App behaves well, simply}.
The \GrtL{} and \AsmL{} around \code{MW.put} and \code{MW.get} explicitly say that we guarantee and assume nothing about them.
The special module-local variable $\mres$ is only used by \AsmL{} and \GrtL{} and initialized with \Run{}.
The intuition behind these conditions is as follows:
$(i)$ only \code{init()} can be invoked initially by an external module since \Run{} is kept in $\mres$ of \App{};
$(ii)$ once \code{init()} is invoked with \Init{}, it consumes \Init{} and returns \Run{} instead, so that \code{init()} cannot be invoked anymore, but instead
$(iii)$ \code{run()} can be invoked with \Run{}, which returns \Run{} back to the caller, so that \code{run()} can be invoked again.

Then, similarly as before, one can informally check that $\appimp \lectx \appmidcond$ holds,
which can be formalized by a simulation relation with the following invariant.
\[
\begin{array}{r@{}l}
  &(\code{initialized}=\code{false} \land \mres = \Run)\\
  {}\lor{} & (\code{initialized}=\code{true} \land \mres = \Init)
\end{array}  
\]

\paragraph{Further abstraction}
In $\appabscond$, we add further conditions about \MW{} to reason about the propagation of \code{42} from \code{init()} to \code{run()}.
For this, we use new resources \code{\MWis{}(f)} for $\code{f} \in \vint\rightarrow\vintopt$, where $\vintopt$ is a shorthand for $\code{option}\;\vint$.
Intuitively, \code{\MWis{}(f)} captures $(i)$ the knowledge that the module \MW{} currently contains the partial map \code{f}
and $(ii)$ the capability to invoke \code{MW.put} and \code{MW.get}.
Technically, these resources are defined using the standard PCM combinators $\authm$ and $\exm$~\cite{irisgroundup} (see \Cref{sec:examples} for the definition of \MWis{}).

Then \code{init()} assumes \Init{} and \code{\MWis{}(fi)} for some \code{fi} at the beginning,
and guarantees \Run{} and \code{\MWis{}($\mapinsert{\code{0}}{\code{42}}{\code{fi}}$)} at the end.
Here it is important to note that the variable \code{fi} connecting the assumption and guarantee, also called \emph{auxiliary variable} in the literature~\cite{auxvar1, auxvar2},
is taken via \kwtake{} making an illusion of receiving the information from the caller.
Similarly, \code{run()} assumes \code{\MWis{}(fi)} with $\code{fi(0)} = \code{Some$\;$42}$ for some \code{fi} together with \Run{} at the beginning,
and guarantees the same at the end.
These conditions form \Asm{\App{} is used well} and \Grt{\App{} behaves well}.

On the other hand, in \code{init()}, we guarantee \code{\MWis{}(fe)} for some \code{fe} before the call \code{MW.put(0,$\;$42)}
and assumes \code{\MWis{}($\mapinsert{\code{0}}{\code{42}}{\code{fe}}$)} after the call.
Similarly, in \code{run()}, we guarantee \code{\MWis{}(fe)} with $\code{fe(0)} = \code{Some$\;$ve}$ for some \code{fe} and \code{ve} before the call \code{v$\;$:=$\;$MW.get(0)}
and assumes \code{\MWis{}(fe)} with $\code{v} = \code{ve}$ after the call.
These form \Grt{\MW{} is used well} and \Asm{\MW{} behaves well}.
Note that this time the auxiliary variables \code{fe} and \code{ve} are chosen via \kwchoose{} making an illusion of sending the information to the callee.

We conclude with a few remarks on the verification of $\appmidcond \lectx \appabscond$.
First, all proof obligations about \Init{} and \Run{} are automatically discharged by the CCR proof mode
using the fact that conditions about them are preserved between $\appmidcond$ and $\appabscond$.
Second, by choosing \code{fe} to be the same as \code{fi} (and \code{ve} to be \code{42}) in \code{init()} and \code{run()},
all the guarantees about \MWis{} are trivially discharged by the immediately preceding assumptions.
Third, from the assumption after the call \code{v$\;$:=$\;$MW.get(0)} in $\appabscond$, it follows that \code{v} is \code{42},
which, together with the fact that the return value \code{v} from the same call should be the same in $\appmidcond$ and $\appabscond$,
proves that the same string is passed to \sysprint{} in \code{run()} of $\appmidcond$ and $\appabscond$.

\section{Formal definitions and key theorems}
\label{sec:formal}

We present 
the underlying semantics \ems,
the specification language \spc with its shallow embedding into \ems,
key theorems of CCR and our verified compiler for \imp{}.

\subsection{\ems{} (Executable Module Semantics)}
\begin{figure}[t]
\small
\begin{minipage}{0.5\textwidth}
$\begin{array}{@{}l@{\ }l@{\ }l@{}}
\myset{X}{cond} &\defeq& \text{if } cond \text{ holds, then } X \text{ else } \emptyset
\\
\trm{fundef}(E) &\defeq& \tyany \rightarrow \code{itree}\;E\;\tyany
\end{array}$
\\
$\begin{array}{@{}l@{\ }c@{\ }l@{}}
\EmsE(X)  &\defeq&
\myset{\curlybracket{\syscallE{} \; \fn \; args \;|\; \fn \in \code{string},\, args \in \tyany}}{X = \tyany} \uplus {}
\\&&
\myset{\curlybracket{\callE \; \fn \; args \;|\; \fn \in \code{string},\, args \in \tyany}}{X=\tyany} \uplus {}
\\&&
\myset{\curlybracket{\getE}}{X=\tyany} \uplus \myset{\curlybracket{\putE\;a \;|\; a \in \tyany}}{X=\unitset} \uplus {}
\\&& 
\{\chooseE \} \uplus \{ \takeE \}
\\
\ems  &\defeq&  \curlybracket{ (\code{init}, \code{funs}) \in \tyany \times
  (\code{string} \fpfn{} \trm{fundef}(\EmsE))}
\end{array}$
$\emsmod \defeq \LD \times (\LD \rightarrow \ems) \qquad \emsmods \defeq \code{list } \emsmod$
\\
$\llink \boilerplate{{}\in \emsmods \rightarrow \emsmods \rightarrow \emsmods} \defeq \code{append}$
\\
$M \lectx M' \boilerplate{{}\in \emsmods} \defeq \forall\, N \in \emsmods\,.\; \beh{M \llink N} \subseteq \beh{M' \llink N} $
\\
$\trm{ObsEvent}\defeq\setofz{(\syscallE{} \; \fn \; args, r) \;|\; \fn \in \code{string},\, args, r \in \tyany}$
\\  
$\begin{array}{@{}l@{\ }c@{\ }l@{}}
\trm{Trace} &\defcoind&
\curlybracket{e :: tr \;|\; e \in \trm{ObsEvent}, tr \in \trm{Trace} } \uplus{}
\\
&&\curlybracket{\trm{Term } v \;|\; v \in \tyany{}} \uplus \curlybracket{\trm{Diverge}} \uplus
\curlybracket{\trm{Error}} \uplus \curlybracket{\trm{Partial}}
\end{array}$
\\
$\text{Beh} \boilerplate{{}\in \emsmods \rightarrow \mathbb{P}\trm{(Trace)}} \defeq ... $
\end{minipage}
\myhrule
\caption{Definitions of Executable Module Semantics (EMS)}
\label{fig:ems}
\end{figure}


\begin{figure*}[t]\small\vspace*{-1mm}
\begin{minipage}{0.49\textwidth}
$\begin{array}{@{}l@{\ }l@{\ }l@{}}
\rProp_\Sigma &\defeq& \Sigma \rightarrow \textbf{Prop} \ \ \ \ \trm{ for } \Sigma \in \trm{PCM}
\\
\trm{Depth} \boilerplate{{}\ni d} &\defeq& \setofz{\otop} \uplus \setofz{\opure{o} \;|\; o \in \ord}
\\
\argo_1 < \argo_2 &\defeq& \exists o_1.~\argo_1\,{=}\,\opure{o_1} \land (\argo_2\,{=}\,\otop \lor \exists o_2.~\argo_2\,{=}\,\opure{o_2} \land o_1 < o_2)
\\
\trm{Cond}_\Sigma \boilerplate{{}\ni s} &\defeq& \setofz{(\Meta, \Dep, \Pre, \Post) \;|\; \Meta \in \mathbf{Set} \land
  \Dep \in \Meta \rightarrow \trm{Depth} \land {} \\
  && \hspace*{3.9pc}\Pre,\Post \in \Meta \rightarrow \tyany \rightarrow \tyany \rightarrow \rProp_\Sigma}
\\
\trm{Conds}_\Sigma \boilerplate{{}\ni S} &\defeq& \code{string} \fpfn{} \trm{Cond}_\Sigma
\\
\SpcE(X) &\defeq& \EmsE(X) \uplus \myset{\curlybracket{\ipcE}}{X=\tyany}
\\
\trm{PAbs} \boilerplate{{}\ni A} &\defeq& \curlybracket{ (\code{init}, \code{funs}) \in
  \tyany \times (\code{string} \fpfn{} \trm{fundef}(\SpcE))}
\\
\toabs{A} \boilerplate{{}\in \ems} &\defeq&  (A.\code{init},\; \lambda\,\fn\,\argv.\; (A.\code{funs} \; \fn\; \argv)[\ipcE{} \mapsto \kwret\;()])
\end{array}$
\\
$\spcmod{S_\trm{in}}{A, \mires}{S_\trm{out}} \boilerplate{{}\in \ems}$
when $\dom(A.\code{funs}) = \dom(S_\trm{out})$
$\defeq$ \\
$\mbox{}\quad((A.\code{init}, \mires),$\\
$\mbox{}\quad\;\lambda\fn\,{\in}\,\dom(A.\code{funs}).\; \ASFun{}(S_\trm{in},\; S_\trm{out}\,\fn,\; A.\code{funs}\,\fn))$
\begin{Verbatim}[commandchars=\\\{\},codes={\catcode`$=3\catcode`_=8},fontsize=\small]
\ASCall{}((\Meta, \Dep, \Pre, \Post), $\argo$, \eres{}, $\fn$, $\argv$) $\defeq$
  $\meta$ \asgn{} \kwchoose(\Meta);
  \kwguarantee($\Dep{}(\meta){}\le{}\argo$);
  ($\argp$, \lres) \asgn{} \Grt{\Pre{}($\meta$), $\argv$, \eres};
  $\retp$ \asgn{} \kwcall $\fn$ $\argp$;
  ($\retv$, \eres) \asgn{} \Asm{\Post{}($\meta$), $\retp$, \lres};
  \kwret ($\retv$, \eres{})
\end{Verbatim}
\begin{Verbatim}[commandchars=\\\{\},codes={\catcode`$=3\catcode`_=8},fontsize=\small]
\color{purple}\Asm{Cond, $\xorvp$, \lres} $\equiv$
\color{purple}  ($\xorvv$,\eres,\res) \asgn{} \kwtake($\tyany\times\Sigma\times\Sigma$);
\color{purple}  (\_, \mresb{}) \asgn{}\downcastU{\tyany\,{\times}\,\Sigma} \kw{get};
\color{purple}  \kwassume(Cond $\xorvp$ $\xorvv$ \res);
\color{purple}  \kwassume$(\mval(\mresb{}+\lres+\eres+\res))$;
\color{purple}  \kwret ($\xorvv$, \eres)
\end{Verbatim}
\end{minipage}%
\vrule$\,${}%
\begin{minipage}{0.5\textwidth}
\begin{Verbatim}[commandchars=\\\{\},codes={\catcode`$=3\catcode`_=8},fontsize=\small]
\ASFun{}($S_\trm{in}$, (\Meta, \Dep, \Pre, \Post), $f$\boilerplate{${}\in{}$\trm{fundef}(\SpcE)}) $\defeq$ $\lambda\,\argp.$
  $\meta$ \asgn{} \kwtake(\Meta);
  ($\argv$, \eres{}) \asgn{} \Asm{\Pre{}($\meta$), $\argp$, $\munit$};
  ($\retv$, \eres{}) \asgn{} \kwmatch \Dep{}($\meta$) \kwwith
    | $\otop$ => $f(\argv)$[
        $\callE{}\,\fn\,\argv$ $\mapsto$ $\lambda\,$\frm{}.$\,$\ASCall{}($(S_\trm{in}\,\fn)\unwrapN{}$, $\otop$, \frm{}, $\fn$, $\argv$),
        \ipcE $\mapsto$ $\lambda\,$\frm{}. \ASIPC{}($S_\trm{in}$, $\otop$, \frm{}),
        \putE \ms{} $\mapsto$ $\lambda\,$\frm{}. (\_, \mresb{}) \asgn{}\downcastU{\tyany\,{\times}\,\Sigma} \kw{get};
                        $\;$\kw{put}\;(\ms{}, \mresb{}); \kwret ((), \frm{}),
        \getE $\mapsto$ $\lambda\,$\frm{}. (\ms{}, \_) \asgn{}\downcastU{\tyany\,{\times}\,\Sigma} \kw{get}; \kwret (\ms{}, \frm{})
      ](\eres{})
    | \opure{o} => \ASIPC{}($S_\trm{in}$, \opure{o}, \eres{}) \kwend;
  ($\retp$, \_) \asgn{} \Grt{\Post{}($\meta$), $\retv$, \eres{}};
  \kwret $\retp$
\end{Verbatim}
\vspace*{-1.5mm}
\begin{Verbatim}[commandchars=\\\{\},codes={\catcode`$=3\catcode`_=8},fontsize=\small]
\ASIPC{}($S_\trm{in}$, $\argo$, \eres{}) $\defeq$ \ASIPCi{}($S_\trm{in}$, $\argo$, \eres{}, \kwchoose(\ord))
\ASIPCi{}($S_\trm{in}$, $\argo$, \eres{}, i) $\defeq$
  \kwif (\kwchoose(bool)) \{
    ($\fn$, $\argv$, $d$') \asgn{} \kwchoose(\fname $\times$ \anyv $\times$ \trm{Depth});
    \kwguarantee($\argo\code{'}<\argo$);
    (\_, \eres{}) \asgn{} \ASCall($(S_\trm{in}\,\fn)\unwrapN{}$, $\argo$', \eres{}, $\fn$, $\argv$);
    \ASIPCi{}($S_\trm{in}$, $\argo$, \eres{}, \kwchoose(\{i' $\in$ \ord | i' < i\}))
  \} \kwelse \{ \kwret (\kwchoose(\tyany), \eres{}) \}
\end{Verbatim}
\vspace*{-1mm}
\begin{Verbatim}[commandchars=\\\{\},codes={\catcode`$=3\catcode`_=8},fontsize=\small]
\color{blue}\Grt{Cond, $\xorvv$, \eres} $\equiv$
\color{blue}  ($\xorvp$,\mresb{},\lres,\res) \asgn{} \kwchoose($\tyany\times\Sigma\times\Sigma\times\Sigma$);
\color{blue}  (\ms{}, \_) \asgn{}\downcastU{\tyany\,{\times}\,\Sigma} \kw{get}; \kw{put}\;(\ms{}, \mresb{});
\color{blue}  \kwguarantee(Cond $\xorvp$ $\xorvv$ \res);
\color{blue}  \kwguarantee$(\mval(\mresb{}+\lres+\eres+\res))$;
\color{blue}  \kwret ($\xorvp$, \lres)
\end{Verbatim}
\end{minipage}%
\hfill\mbox{}\vspace*{-1mm}
\myhrule
\caption{Definition of \spc and its shallow-embedding into \ems.}
\label{fig:spc}
\end{figure*}


\paragraph{Interaction trees}
First of all, our Coq formalization largely relies on interaction trees~\cite{liyao:itree}.
Intuitively, an \emph{itree} in $\code{itree}\;E\;T$ for an event type $E: \mathbf{Set}\rightarrow\mathbf{Set}$ (consisting of a set of events $E(X)$ whose return type is $X$)
and a return type $T$ for the itree
can be understood as an open small-step operational semantics that can $(i)$ take a silent step, $(ii)$ terminate with a return value of type $T$,
or $(iii)$ trigger an event in $E(X)$ for some $X$
and, as a continuation, give an itree for each possible return value in $X$.
We enjoy two benefits of interaction trees:
$(i)$ they are extracted to executable programs in OCaml, and
$(ii)$ they provide useful combinators, which made our various constructions straightforward.

We mainly use the \emph{interpretation} combinator with the type:
\[
\begin{array}{@{}l@{}r@{}}
  \code{itree}\;E\;T &{}\rightarrow (\forall X.\; E(X) \rightarrow ST \rightarrow \code{itree}\;E'\;(X\times ST))\\
  & {}\rightarrow ST \rightarrow \code{itree}\;E'\;(T\times ST)\ \,{}
\end{array}
\]
It takes an itree $t \in \code{itree}\;E\;T$, adds a local state of type $ST$,
and interprets each event in $E$ as an itree in a new event type $E'$ that can access and update the local state.
This combinator is useful when adding the code encoding conditions (\ie those written in \asm{red} and \grt{blue} in the previous examples)
by interpreting each call event as the same call with conditions around it.
We use the notation $t[e_1 \,{\mapsto}\, \lambda s.\, t_1,\ldots,e_n \,{\mapsto}\, \lambda s.\,t_n](s_0)$
to denote the resulting itree
when the combinator is applied to an itree $t$, with an initial local state $s_0$, by interpreting each event $e_i$ to an itree $t_i$ for a given local state $s$.
We omit those events that are interpreted identically,
and the state component when it is the unit type.
Since $\code{itree}\;E$ forms a monad for any $E$,
we henceforth use the monad notations: $x \,\asgn{}\, t_1; t_2$ for bind and $\kwret{}\,v$ for return.


\paragraph{\ems}
\Cref{fig:ems} shows the formal definition of \ems,
where we use $\myset{X}{cond}$ to denote a conditionally non-empty set.
First, $\trm{fundef}(E)$ is the semantic domain for a function,
which takes a value in \tyany{} as an argument and
gives an itree w.r.t. the event type $E$ and the return type \tyany{},
where \tyany{} can be understood as the set of all mathematical values.
$\EmsE$ is the event type for \ems consisting of
$(i)$ $\syscallE$ for triggering observable events such as system calls,
$(ii)$ $\callE$ for making a call to (internal or external) functions,
$(iii)$ $\getE$ and $\putE$ for accessing the module local state of type \tyany{},
and $(iv)$ $\chooseE$ and $\takeE$ for nondeterministically \emph{choosing} and \emph{taking} a value from any given set $X$.
\ems{} is the semantic domain for a module (after loading),
which is given by $(i)$ the initial value of the module local state, \code{init}, and
$(ii)$ the definitions of the module's functions, \code{funs}, with the event type $\EmsE$.

$\emsmod$ gives a notion of module code (\ie before loading) for a global loading data type $\LD$,
which happens to be required to form a PCM
to combine loading data from all modules and express consistency between them.
A \emph{module code} consists of its own loading data in $\LD$ and a loading function in $\LD \rightarrow \ems$
that, given the global loading data gathered from all the modules, returns its module semantics.
A \emph{modules code} in $\emsmods$ is simply a list of module codes
and linking $\link$ between them is the list append.
Note that we require function names to be \emph{globally} unique when loading modules,
which is not too strong since function names can include their module name as a prefix.
Then we define contextual refinement between two modules codes $M$ and $M'$ as behavioral refinement under an arbitrary context modules code $N$.

\paragraph{Observable behavior}
To give the notion of behavior, we first define the set of traces, Trace, coinductively.
A trace is a finite or infinite sequence of \trm{ObsEvent}
(\ie pairs of an observable event and its return value)
that can possibly end with one of the four cases:
$(i)$ normal termination with an \anyty{} value,
$(ii)$ silent divergence without producing any events,
$(iii)$ erroneous termination,
or $(iv)$ partial termination.
The notion of \emph{partial termination} is interesting, which is used to define \grt{NB} (to be shown below).
It can be intuitively understood as stopping the execution at the user's will
such as pressing \texttt{Ctrl+C},
which is dual to erroneous termination (\ie termination due to the program's fault).

The predicate $\textrm{Beh}(M)$ defines all possible traces of the modules code $M$
in a standard way except for the following.
First, the partial termination, \textrm{Partial}, can occur nondeterministically at any point during execution
(capturing that the user can stop the program at any time).
Second, the behaviors of \kwchoose{} and \kwtake{} are defined as follows:
\[
\begin{array}{@{}l@{}c@{}l@{}}
  \behems(\code{x$\,$\asgn{}$\,$\kwchoose($X$);$\,${}$K[\code{x}]$}) &\defcoind& \setofz{\trm{Partial}} \cup \bigcup_{x\in X} \behems(K[x]) \\
  \behems(\code{x$\,$\asgn{}$\,$\kwtake($X$);$\,${}$K[\code{x}]$}) &\defcoind& \setofz{\trm{Partial}} \cup \bigcap_{x\in X} \behems(K[x]) \\
\end{array}
\]
where $\behems$ is coinductively defined for the large (closed) itree that
is obtained after loading by combining all the itrees of all functions of all modules in $M$ (see \cite[Fig.~9-10]{supplementary} for definition).
Third, \kwassume{} and \kwguarantee{} are defined as follows.
\smallskip\\
{\mbox{}\hfill$
\begin{array}{@{}l@{\ \ }c@{\ \ }l@{}}
  \kwassume(P) &\defeq& \kwif\;(P)\;\kwskip\;\kwelse\;\kwtake(\emptyset) \\
  \kwguarantee(P) &\defeq& \kwif\;(P)\;\kwskip\;\kwelse\;\kwchoose(\emptyset) \\
\end{array}
$\hfill\mbox{}}
\smallskip\\
Note that \code{\kwassume(False)} exhibits all observable behaviors (\ie \trm{Trace}), understood as \asm{UB},
and \code{\kwguarantee(False)} exhibits only the partial termination (\ie $\setofz{\trm{Partial}}$), understood as \grt{NB}.
Thanks to the partial termination, we can avoid the completely empty behavior, which may cause a trouble since it can eliminate previously triggered events.

\subsection{\spc{} and its embedding into \ems{}}
\label{sec:formal:spc}

We define a language, \spc{}, where one can specify conditions and write executable code for modules,
which are then shallowly embedded into \ems (\ie auto-generating the code that encodes the conditions).
\Cref{fig:spc} shows the formal definition of \spc{} and its embedding.

\paragraph{\spc{}}
In \spc{}, for each function, we can specify a pair of pre and post conditions $s \in \trm{Cond}_\Sigma$, which is parameterized by a global PCM $\Sigma$.
Concretely, a \emph{condition} $s \in \trm{Cond}_\Sigma$ consists of four components $(\Meta, \Dep, \Pre, \Post)$
and a \emph{collection of conditions} $S \in \trm{Conds}_\Sigma$ consists of such conditions for a finite set of functions.
Here $\Meta$ defines the type of the auxiliary variable~$\meta$ (\eg \code{fi} and \code{fe} in \Cref{fig:app-def}) that is shared among \Dep{}, \Pre{} and \Post{},
which are explained below.


$\Dep(\meta)$, given $\meta \in \Meta$, specifies the maximum call depth.
This component is used to specify \emph{pure calls}, which can be automatically eliminated by the ACT theorem
since they $(i)$ always terminate $(ii)$ without triggering any observable event.
For example, in \Cref{fig:mw-def}, calls to \Mem{} and \Map{} are all eliminated
in $\mwabs$ since they are pure.
Specifically, a depth $d\in\trm{Depth}$ is either $\otop$ denoting (potential) impurity, or an ordinal $\opure{o}$ denoting purity (with a maximum call depth $o$).
We then give a well-founded ordering~$<$ on $\trm{Depth}$.

Now we see how we can \emph{locally} impose purity (\ie termination and absence of observable events).
First,
we require the depth to \emph{strictly} decrease for a pure call (see line 3 in \ASIPCi{}).
From this (at the point of applying ACT) it follows that any chain of pure calls always terminates.
Second,
when a function is invoked with depth $\opure{o}$ (\ie a pure call),
its body is replaced with \ASIPC{} that can only nondeterministically make a finite number of arbitrary pure calls
(see line 11 in \trm{FunDef}).
From this (at the point of applying ACT) it follows by construction that pure calls do not trigger any observable event.
Note that purity of a function may depend on its argument (\eg printing an error for an invalid argument but otherwise behaving purely).
The 
technical report~\cite[\S3.2]{supplementary} presents more detailed explanation about purity with an example.

$\Pre(\meta)$/$\Post(\meta)$, given $\meta \in \Meta$, specifies a pre/post condition on $(i)$ a \emph{concrete} argument/return value, $(ii)$ an \emph{abstract} argument/return value, and $(iii)$ an argument/return resource.
The notions of concrete and abstract values are used to abstract values passed between functions.
For example, we can abstract a function taking a pointer to a linked list (\ie concrete value) into that taking a mathematical list (\ie an abstract value).
The 
technical report~\cite[\S3.4]{supplementary} presents such an example.
%

It is important to note that even though it does not make sense to send or receive abstract values to and from contexts since the contexts are arbitrary,
we can again make such an illusion using \kwchoose{} and \kwtake{}.
Specifically, when an abstract function passes an abstract value to a context, we \emph{choose} a concrete value that satisfies the required
pre/post condition together with the abstract value and the chosen resource; conversely, when a context passes a concrete value,
we \emph{take} an abstract value satisfying the required condition and pass it to the abstract function.

Note also that \spc allows us to specify conditions as $\rProp_\Sigma$-level predicates following \iris~\cite{irisgroundup} and
supports the IPM (\iris Proof Mode)~\cite{krebbers:proofmode} for reasoning about them.

In \spc{}, we can also write abstract yet executable code for each function as an itree,
a collection of which form a \emph{pre-abstraction} $A \in \trm{PAbs}$.
Concretely, $A$ is the same as an \ems module except that it can trigger an extra event, \ipcE{},
which is interpreted as nondeterministically making arbitrary pure calls (w.r.t. the input conditions) in conditional abstractions but eliminated after applying the ACT theorem.
Note that 
\ipcE{} is implicitly inserted at each line (via a macro expansion for the bind operator)
because they will be freely eliminated by ACT.
For example, in \mbox{\Cref{fig:mw-def}},
when \code{Mem.store} or \code{Mem.load} is invoked in $\mwimp{1}$,
the same call can be made in $\mwmidcond$ via \ipcE{} since they are pure calls.

\paragraph{Embedding into \ems{}}
A pre-abstraction $A$ together with conditions are translated into \ems in two ways as follows.
\begin{itemize}[leftmargin=8pt]
\item $\toabs{A} \in \ems$, called \emph{abstraction}, is obtained by eliminating all \ipcE{} events (\ie replacing them by $\kwret\;()$).
\item $\spcmod{S_\trm{in}}{A, \mires}{S_\trm{out}} \in \ems$, called \emph{conditional \mbox{abstraction}},
has $(A.\code{init}, \mires)$ as an initial module state together with function definitions generated by \ASFun{},
where
$S_\trm{in}$ is conditions about the functions that $A$ invokes, $S_\trm{out}$ conditions about the functions that $A$ defines, and $\mires$ an initial module resource of $A$.
\end{itemize}
For example, $\mwmid$, $\mwabs$, $\appmid$ and $\appabs$ (with implicit \ipcE{}'s) are (pre-)abstractions and $\mwmidcond$, $\mwabscond$, $\appmidcond$ and $\appabscond$ are conditional abstractions.

\originaltxt{
A \emph{pre-abstraction} $A \in \trm{PAbs}$ is an abstract module that is translated into \ems in two ways.
Specifically, $A$ is the same as an \ems module except that it can trigger an extra event, \ipcE{},
and translated as follows.
\todo{which is interpreted as arbitrary pure calls in conditional abstractions but eliminated after applying the ACT theorem.}
\todo{For example, ...}
}



\ASFun{} is a formal definition of what we have explained so far,
where by using the interpretation function for itrees,
we introduce the \frm{} variable and replace $(i)$ each call with \ASCall{}, which adds the condition for the callee around the call; $(ii)$ each \ipcE{} with \ASIPC{}, which makes a finite number of arbitrary pure calls, and $(iii)$ \putE{} and \getE{} with accessing the first component of the module local state since it is extended with \mresb{} in the second component.
Note that $(S_\trm{in}\,\fn)\unwrapN{}$ triggers \grt{NB} when $\fn \not\in \dom(S_\trm{in})$,
and $\asgn{}\downcastU{\tyany\,{\times}\,\Sigma}$ triggers \asm{UB} when the assigned value is not of type $\tyany\,{\times}\,\Sigma$.

\begin{figure*}[t]\small
\begin{minipage}[t]{0.5\textwidth}
$\mires_\Mem \trm{ := } \authfull\munit \in \authm{\,(\vptr \rightarrow \exm{\,(\vval)})} \subseteq \Sigma$
\\
$\begin{array}{@{}l@{}}
S_\Mem := \{\\
\code{\Mem.alloc}{:}\, \forall (\argo,n):\trm{Depth}\times\vint.~\curlybracket{\argo}
\\
\;\;\curlybracket{\lambda\,\argp.\; \rplift{\argo=\otop}\lor\rplift{x = {[n]} \land n \ge 0} }
\\
\;\;\curlybracket{\lambda\,\retp.\; \rplift{\argo=\otop}\lor\exists\,p{:}\vptr, \lst{:}\vlist\;\vval.\, {(p \pointsto \lst)} * \rplift{\!\retp\,{=}\,{p} \land \lstlength(\lst) \,{=}\, n\!} },
\\
\code{\Mem.free}{:}\, \forall \argo:\trm{Depth}.~\curlybracket{\argo}
\\
\;\;\curlybracket{\lambda\,\argp.\;\rplift{\argo=\otop}\lor \exists\,p\,{:}\,\vptr.\; (p \pointsto [\_]) * \rplift{\argp = {[p]}} }
\\
\;\;\curlybracket{\lambda\,\retp.\;\rplift{\argo=\otop}\lor \rplift{\retp \in {\vval}}},
\\
\code{\Mem.load}{:}\, \forall (\argo,p,v): \trm{Depth}\times\vptr\times\vval.~\curlybracket{\argo}
\\
\;\;\curlybracket{\lambda\,\argp.\;\rplift{\argo=\otop}\lor ((p \pointsto [v]) * \rplift{\argp = {[p]}}) }
\\
\;\;\curlybracket{\lambda\,\retp.\;\rplift{\argo=\otop}\lor ((p \pointsto [v]) * \rplift{ \retp = {v}}) },
\\
\code{\Mem.store}{:}\, \forall (\argo,p,v): \trm{Depth}\times\vptr\times\vval.~\curlybracket{\argo}
\\
\;\;\curlybracket{\lambda\,\argp.\;\rplift{\argo=\otop}\lor ((p \pointsto [\_]) * \rplift{\argp = {[p, v]}}) }
\\
\;\;\curlybracket{\lambda\,\retp.\;\rplift{\argo=\otop}\lor ((p \pointsto [v]) * \rplift{\retp \in {\vval}}) }
\quad\}
\end{array}$
\\
\hrule\mbox{}\\
$\begin{array}{@{}l@{}}
\mires_\App \trm{ := } \Run \in \appra \subseteq \Sigma
\\  
S_\App := \{
\\
\code{\App.init}{:}\,
\forall f:\vint \rightarrow \vintopt.~\curlybracket{\otop} \\
\;\;\curlybracket{\lambda\,\argp.\; \ownGhost{}{\Init} * \ownGhost{}{\MWis(f)} * \rplift{\argp = {[]}} }
\\
\;\;\curlybracket{\lambda\,\retp.\; \ownGhost{}{\Run} * \ownGhost{}{\MWis(\mapinsert{0}{42}{f})} * \rplift{\retp \in \vval} },
\\[.5mm]
\code{\App.run}{:}\,
\forall f:\vint \rightarrow \vintopt.~\curlybracket{\otop} \\
\;\;\curlybracket{\lambda\,\argp.\; \ownGhost{}{\Run} * \ownGhost{}{\MWis(f)} * \rplift{\argp = {[]} \land f(0) = \some\;42} }
\\
\;\;\curlybracket{\lambda\,\retp.\; \ownGhost{}{\Run} * \ownGhost{}{\MWis(f)} * \rplift{\retp \in \vval \land f(0) = \some\;42} }
\quad\}
\end{array}$
\end{minipage}%
\vrule%
\begin{minipage}[t]{0.5\textwidth}
$\;\mires_\Map \trm{ := } \munit \in \Sigma$
\\  
${}\;\begin{array}{@{}l@{}}
S_\Map := \{ \\
\code{\Map.new}{:}\, \forall \_:\unitset.~\curlybracket{\opure{1}}\\
\;\;\curlybracket{\lambda\,\argp.\; \rplift{\argp = {[]}}}
\\
\;\;\curlybracket{\lambda\,\retp.\; \exists\, h\,{:}\,\vptr.\; \ownGhostMap{}{h\;\contains (\lambda\_.\;\none)} * \rplift{\retp = {h}} },
\\
\code{\Map.update}{:}\, \forall (h,f,k,v):\vptr {\times} (\vint {\rightarrow} \vintopt) {\times} \vint {\times} \vint.~ \curlybracket{\opure{1}}\\
\;\;\curlybracket{\lambda\,\argp.\; \ownGhostMap{}{h\;\contains f} * \rplift{\argp = {[h, k, v]}} }
\\
\;\;\curlybracket{\lambda\,\retp.\; \ownGhostMap{}{h\;\contains \mapinsert{k}{v}{f}} * \rplift{\retp \in \vval} },
\\
\code{\Map.get}{:}\, \forall (h,f,k,v):\vptr \times (\vint {\rightarrow} \vintopt) \times \vint \times \vint.~\curlybracket{\opure{\omega}} \\
\;\;\curlybracket{\lambda\,\argp.\; \ownGhostMap{}{h\;\contains f} * \rplift{\argp = {[h, k]} \land f(k) = \some\;v} }
\\
\;\;\curlybracket{\lambda\,\retp.\; \ownGhostMap{}{h\;\contains f} * \rplift{\retp = {v}} } 
\quad\}
\end{array}$
\\
\hrule\mbox{}\\
$\mbox{}\;\begin{array}{@{}l@{}}
\mires_\MW \text{ := } \authfull\exinj(\lambda\,\_.\;\none) \in \authm{\,(\exm{\,(\vint \rightarrow \vintopt)})} \times \simplera \subseteq \Sigma
\\    
S_\MW := \{
\\
\code{\MW.main}{:}\,
\forall \_:\unitset.~\curlybracket{\otop} \\
\;\;\curlybracket{\lambda\,\argp.\; \ownGhost{}{\Init} * \ownGhost{}{\MWis(\lambda\,\_.\;\none)} * \rplift{\argp = {[]}} }
\\
\;\;\curlybracket{\lambda\,\retp.\; \rplift{\retp \in \vval} },
\\
\code{\MW.put}{:}\,
\forall (f,k,v):(\vint \rightarrow \vintopt) \times \vint \times \vint.~\curlybracket{\otop} \\
\;\;\curlybracket{\lambda\,\argp.\; \ownGhost{}{\MWis(f)} * \rplift{\argp = {[k, v]}} }
\\
\;\;\curlybracket{\lambda\,\retp.\; \ownGhost{}{\MWis(\mapinsert{k}{v}{f}} * \rplift{\retp \in \vval} },
\\
\code{\MW.get}{:}\,
\forall (f,k,v):(\vint \rightarrow \vintopt) \times \vint \times \vint.~\curlybracket{\otop}\\
\;\;\curlybracket{\lambda\,\argp.\; \ownGhost{}{\MWis(f)} * \rplift{\argp = {[k]} \land f(k) = \some\;v} }
\\
\;\;\curlybracket{\lambda\,\retp.\; \ownGhost{}{\MWis(f)} * \rplift{\retp = {v}} }
\quad\}
\end{array}$
\end{minipage}\vspace*{-.7mm}
\myhrule
\caption{Conditions for \Mem{}, \Map{}, \MW{}, \App{} written in \spc{}}
\label{fig:specs}
\end{figure*}


\subsection{Key theorems of CCR}

\begin{theorem}[Assumption Cancellation Theorem (ACT)]\label{thm:ACT}
  For a global PCM $\Sigma$,
  conditional abstractions $\spcmod{S}{(A_i, \mires_i)}{S_i}$ for $i\in\{1,\text{.}\text{.}\text{.}\,, n\}$ with $S \sqsubseteq S_1 \cup \ldots \cup S_n$,
  and an initial resource $\mires$ to $\main$ satisfying its precondition and
  \code{$\mval(\mires$\,+\,$\mires_1$\,+\,$\ldots$\,+\,$\mires_n)$},
\[
\spcmod{S}{(A_1, \mires_1)}{S_1} \link \ldots \link \spcmod{S}{(A_n, \mires_n)}{S_n}
\lectx
\toabs{A_1} \link \ldots \link \toabs{A_n}
\]
\end{theorem}
\noindent
Here the relation $\sqsubseteq$ on $\trm{Conds}_\Sigma$ generalizes the simple inclusion relation following \iris~\cite{irisgroundup} (see the 
technical report~\cite[Fig.~12]{supplementary} for the formal definition).
\hide{ 
\[\small
\begin{array}{@{}l@{\ }l@{\ }l@{}}
P \vdash Q &\defeq& \forall a.~ (\wdef(a) \land P\,a) \implies Q\,a
\\
\upd{} P &\defeq& \lambda a.~ \forall c.~ \wdef(a \madd c) \implies \exists b.~ \wdef(b \madd c) \land P\,b
\\
s_{0} \sqsubseteq s_{1} &\defeq&
\forall \meta_{0} \in (s_{0}.\Meta).\;
\exists \meta_{1} \in (s_{1}.\Meta).\; s_{1}.\Dep{}(\meta_{1}) \sqsubseteq s_{0}.\Dep{}(\meta_{0})
\land{}
\\
&&(\forall\,\argp\,\argv.\; s_{0}.\Pre{}(\meta_{0})\,\argp\,\argv \vdash \upd{} s_{1}.\Pre{}(\meta_{1})\,\argp\,\argv)
\land{}
\\
&&(\forall\,\retp\,\retv.\; s_{1}.\Post{}(\meta_{1})\,\retp\,\retv \vdash \upd{} s_{0}.\Post{}(\meta_{0})\,\retp\,\retv)
\\
S_{0} \sqsubseteq S_{1} &\defeq& \forall \fn, s_{0}.~
S_{0}(\fn) = s_{0} \implies
\exists s_{1}.~ S_{1}(\fn) = s_{1} \land s_{0} \sqsubseteq s_{1}
\end{array}
\]
}




We remark that as a corollary of the ACT theorem, CCR can also serve as a framework for modern separation logics, 
but in an operational style without step-indexing.
For this, we define special \mbox{pre-abstractions} $\Safe(\names_\trm{in}, \names_\trm{out})$,
which defines functions with their names in $\names_\trm{out}$ to only nondeterministically invoke
arbitrary functions in $\names_\trm{in}$ with arbitrary arguments
for any (finite or infinite) number of times.
\begin{lemma} [Safety]\label{thm:erasure-safe}
  For $\names \subseteq \names_{1} \uplus \ldots \uplus \names_{n}$,
\[
  \Safe(\names, \names_{1}) \llink \ldots \llink \Safe(\names, \names_{n})
  \text{ does not produce an error.}
\]
\end{lemma}
\begin{corollary} [SL]\label{thm:erasure-subsume}
  Given a global PCM $\Sigma$,
  $(P_{i}, S_{i}, \mires_{i})$ for $i\in\{1,\text{.}\text{.}\text{.}\,, n\}$
  with $S \sqsubseteq S_1 \cup \ldots \cup S_n$,
  and an initial resource $\mires$ to $\main$ satisfying its precondition and \code{$\mval(\mires$\,+\,$\mires_1$\,+\,$\ldots$\,+\,$\mires_n)$},
\[
  \begin{array}{@{\ \ }ll@{}}
    & (\forall i.~ P_{i} \lectx \spcmod{S}{\Safe(\trm{dom}(S),\trm{dom}(S_{i})), \mires_{i}}{S_{i}})
    \\
      {}\implies & P_{1}{} \llink \ldots \llink P_{n} \text{ does not produce an error.}
  \end{array}
\]
\end{corollary}
This corollary can be seen as a separation logic because
$P_{i} \lectx \spcmod{S}{\Safe(\trm{dom}(S),\trm{dom}(S_{i})), \mires_{i}}{S_{i}}$
essentially amounts to proving, in SL, that $P_{i}$ satisfies the pre and post conditions of $S_{i}$ assuming other modules satisfy those of $S$.
Moreover, by employing the \iris Proof Mode~\cite{krebbers:proofmode}, the actual proofs of these refinements in CCR look similar to those in \iris.


\originaltxt{
Now we present our core theorems. The most important one -- spec erasure theorem -- is
already presented in \Cref{thm:spec-erasure}. \yj{we didn't give general theorem yet}

\todo{Say the theorem for open programs}

\todo{Say the safety as a corollary}

Then we present adequacy of the simulation relation.
\begin{theorem} [Adequacy]\label{thm:adequacy} For a given pair of module $M_\iside$ and $M_\aside$,
  a possible world $\world$ equipped with $\simle$, and a module-local relational invariant $\simrel$,
  if each pair of functions with the same name for $M_\iside$ and $M_\aside$
  are related by the simulation relation $\lesssim$
  for any argument and any module-local states satisfying $I$,
  then we have the contextual refinement between them:
\[
  [M_\iside] \lectx [M_\aside]
\]
\end{theorem}

We also use the following strengthening theorem.
\begin{theorem} [Strengthening]\label{thm:weakening}
  For any $S, S', A, \mires, S_A$, the following holds:
\[
S' \sqsupseteq S \implies \spcmod{S}{A, \mires}{S_A} \lectx \spcmod{S'}{A,\mires}{S_A}
\]
\end{theorem}

Finally, we briefly discuss how to define the abstraction \Safe mentioned in \Cref{sec:introduction}.
The module $\Safe(\names, \names')$ defines each function in $\names'$ whose
semantics is simply defined to nondeterministically invoke an arbitrary function in $\names$ with arbitrary arguments any number of times (even infinitely many).
Then we have the following theorem.
\begin{theorem} [Safety]\label{thm:erasure-safe}
  For $\names = \names_{1} \uplus \ldots \uplus \names_{n}$, $\Safe(\names, \names_{1}) \llink \ldots \llink \Safe(\names, \names_{n})$ is safe
  (\ie never produces an \trm{Error}).
\end{theorem}
}


\subsection{Imp and its verified compiler}\label{sec:imp}

The \imp{} language, extended from \code{Imp}\cite{liyao:itree}, has standard syntax and semantics built on a simplified version of the CompCert memory model.
In particular, \imp{} computes with the set of values, \vval{}, consisting of 64-bit integers, \vint{}, and memory and function pointers, \vptr{}.
When embedding \imp{} into \ems{}, we cast back and forth between \vval{} and \tyany{},
and if the downcast from \tyany{} to \vval{} fails, trigger \asm{UB}.

We also develop a verified compiler\footnote{
  As a simple solution to resolve a subtle mismatch between CompCert's memory model and ours, 
  we compile the \code{free} instruction to \code{skip} for now.
  Also we support separate compilation following the approach of \cite{kang:scc}.
}
from \imp to Csharpminor of CompCert~\cite{CompCert},
which is then composed with CompCert to give a verified compiler $\impcomp{-}$ from \imp{} to 
assembly.

\begin{theorem} [Separate Compilation Correctness]\label{imp:correctness}
  Given $(P_i,Asm_i)$ with $\impcomp{P_i} = \some\;Asm_i$
  for $i\in\{1,\text{.}\text{.}\text{.}\,, n\}$,\\
$
  \beh{Asm_1 \plink \cdots \plink Asm_n}\footnote{
    We cast CompCert's events into \code{Obs} events in \ems.
  }
  \subseteq \beh{\memimp \llink P_1 \llink \cdots \llink P_n}
$
\end{theorem}
\noindent
Here $\plink$ is the \emph{syntactic linking} operator of CompCert, and 
$\memimp$ is an \ems{} module (directly written as itrees) that
implements our memory model (\ie a simplified version of CompCert's).




\section{Examples}
\label{sec:examples}




\Cref{fig:specs} shows the conditions for the modules used in the motivating example of \Cref{fig:mw-def}, where we use the notation
\mbox{$\forall \meta: \Meta.~\curlybracket{D}\curlybracket{\lambda\,\argp.~P}\curlybracket{\lambda\,\retp.~Q}$}
as a shorthand for the condition
$(\Meta, (\lambda\meta.~D), (\lambda\meta\,\argp\,\argv.~\rplift{\argp = \argv} * P),
(\lambda\meta\,\retp\,\retv.~\rplift{\retp = \retv} * Q)) \in \trm{Cond}_\Sigma$
for $P,Q \in \rProp_\Sigma$,
and use standard \iris{} notations and PCMs~\cite{irisgroundup}:
$\rplift{}$ for lifting \textbf{Prop} to $\rProp_\Sigma$,
$\ownGhost{}{}$ for owning a resource, $*$ for separating conjunction, and
\exm(\_) and \authm(\_) for exclusive and authoritative PCMs.


For \Mem{}, we use the standard ``points-to'' resources ${p \pointsto l}$ 
and its abstraction $\memabs$ is defined the same as its implementation $\memimp$
because $(i)$ when a function of \Mem{} is invoked impurely with $\otop$,
both pre and post conditions amount to \code{True} and its abstraction simply becomes its implementation; and
$(ii)$ when invoked purely with $\opure{o}$,
the pre and post conditions amount to the standard ones in SL and its abstraction is ignored and replaced by \ASIPC{}.
Therefore, in an incremental verification, we can abstract a selected set of memory operations into \ipcE{}
leaving the rest as unabstracted, which can be abstracted later in a subsequent verification.

For \Map{}, $\mapimp$ implements maps as linked lists using \Mem{}
and $\mapabs$ is empty (\ie triggering \grt{NB} when invoked)
since all its functions are pure and thus every call to them is eliminated by ACT.
Its conditions use standard resources for a key-value linked list, $h\;\contains f$,
defined as 
$(\exists \; hd \; \ell.~ {(h \pointsto hd)} * \text{is\_list} \; hd \; \ell * \rplift{f = \lambda k.~ \ell \;!!\; k})$.
For \MW{} and \App{},
$\MWis{(f)}$ denotes $(\authfrag(\exinj(f)) + \half{})$ where $\half{} \in \simplera{}$ represents mere capability to invoke functions of $\MW{}$,
which is used in $\mwmidcond$ as a part of ``\MW{} is used/behaves well, simply'' to prevent \Mem{} from calling them by not passing $\half{}$ to \Mem{}.



\section{Reasoning about function pointers}
\label{sec:advanced-higher}
\begin{figure}[t]\small
$P_\RP$ :=
\begin{minipage}[t]{0.4\textwidth}
\begin{Verbatim}[commandchars=\\\{\},codes={\catcode`$=3},fontsize=\small]
[\kwmodule{} RP]
\kwdef repeat(f:\vptr{}, n:\vint{}, m:\vint{}) $\equiv$
  \kwif n $\leq$ 0 \kwthen \kwreturn m
  \kwelse \{ \kwvar v := (*f)(m)
         \kwreturn RP.repeat(f, n-1, v) \}
\end{Verbatim}
\end{minipage}%
\hfill\mbox{}
\\[1mm]
$P_\SC$ :=
\begin{minipage}[t]{0.4\textwidth}  
\begin{Verbatim}[commandchars=\\\{\},codes={\catcode`$=3},fontsize=\small]
[\kwmodule{} SC] \kwdef succ(m:\vint{}) $\equiv$ m + 1
\end{Verbatim}
\end{minipage}
\hfill\mbox{}\\[1mm]
\hrule\mbox{}
\begin{minipage}[t]{0.315\textwidth}
$P_\AD$ := \code{[\kwmodule{} AD]}
\begin{Verbatim}[commandchars=\\\{\},codes={\catcode`$=3},fontsize=\small]
\kwdef main() $\equiv$
  \kwvar n := \sysgetint()  
  \sysprint(str(RP.repeat(&SC.succ,n,n)$\!$)$\!$)
\end{Verbatim}
\end{minipage}\vrule$\;$%
\begin{minipage}[t]{0.25\textwidth}
$A_\AD$ := \code{[\kwmodule{} AD]}
\begin{Verbatim}[commandchars=\\\{\},codes={\catcode`$=3},fontsize=\small]
\kwdef main() $\equiv$
  \kwvar n := \sysgetint()  
  \sysprint(str(n + n)$\!$)
\end{Verbatim}
\end{minipage}%
\\[1mm]
\hrule\mbox{}
$\begin{array}{@{}l@{}}
H_\RP(\Sf) \!:=\! \{ \\
\ \ \code{RP.repeat}: \forall (f, n, m, \fsem):\vptr{\times}\vint{\times}\vint{\times}(\vint{\rightarrow}\vint).\;
\\
\quad\curlybracket{\opure{\omega+n}}
\\
\quad\{ \lambda\,\argp.\; \rplift{\argp = {[f, n, m]} \lland n \ge 0 \lland{}
\\
\quad\quad
\Sf \sqsupseteq \{ \code{*}\!f{:}\, \forall m\!:\!\vint, \curlybracket{\omega} \curlybracket{\lambda\,\argp.\, \rplift{\!\argp \,{=}\, {[m]}}}\curlybracket{\lambda\,\retp.\, \rplift{\!\retp \,{=}\, \fsem(m)\!}}\!\}\!}\}
\\
\quad \curlybracket{\lambda\,\retp.\; \rplift{\retp = {{\fsem}^n(m)}}} \}
\\  
S_\SC{:=}\{ \code{SC.succ}{:}\,\forall m\,{:}\,\vint.\, \curlybracket{\opure{0}} \curlybracket{\lambda\,\argp.\, \rplift{\argp \,{=}\, {[m]}}} \curlybracket{\lambda\,\retp.\, \rplift{\retp\,{=}\, m\,{+}\, 1}}\!\}
\\
S_\AD{:=}\{ \code{AD.main}{:}\,\forall \_\,{:}\,().\, \curlybracket{\otop}\curlybracket{\lambda\,\argp.\; \rplift{\argp=[]}}\curlybracket{\lambda\,\retp.\; \rplift{\retp \in \vval{}}} \}
\end{array}$\vspace*{-1.8mm}
\myhrule
\caption{An example of higher-order reasoning}
\label{fig:repeat}
\end{figure}





We present a general pattern for doing higher-order reasoning in CCR without requiring any special support.
For this, consider the simple example given in \Cref{fig:repeat}.
The function $\code{repeat(f,n,m)}$ in $P_\RP$ recursively apply \code{*f}, \code{n} times, to \code{m},
where \code{*f} is the function pointed to by the pointer value \code{f}.
The definitions in $P_\SC$ and $P_\AD$ are straightforward to understand except that
\code{\&\SC.succ} is the pointer value pointing to the function \code{\SC.succ}.
The pre-abstractions $A_\RP$ and $A_\SC$ are empty since they are pure.
The pre-abstraction $A_\AD$ turns the call to \code{\RP.repeat} into the addition.

To specify \code{\RP.repeat}, we essentially need to embed expected conditions for argument functions \code{f} inside the condition of \code{\RP.repeat}.
Directly supporting this would make the definition of condition more involved since we need to solve a recursive equation to define it.
Although such an equation could be solved by employing the step-indexing technique,
here we propose a more elementary solution that does not introduce any cyclic definition.

Now we see how to do it.
First, we give a higher-order condition $H_\RP$ to the module $\RP$, given in \Cref{fig:repeat},
which is given as a function from conditions to conditions.
Concretely,
given $\Sf$, 
for arguments $f,n,m$ and a mathematical function $\fsem$,
the condition $H_\RP(\Sf)$ assumes $\Sf$ to include the expected specification for \code{*$f$}
(saying that \code{*$f$} is pure with measure $\opure{\omega}$ and returns $\fsem(m)$ for any argument $m$),
and then guarantees that the return value is ${\fsem}^n(m)$.
Here $\omega$ is the smallest ordinal bigger than every natural number and thus \code{*$f$} is allowed to have any \emph{finite} recursion depth.
Also we require \code{\RP.repeat} to be pure with measure 
$\opure{\omega+n}$
because it makes recursive calls with depth $n$ followed by \mbox{a call to \code{*$f$}}.

Then we verify $\RP$. For any $\Sf$ and any $S \sqsupseteq (\Sf \cup H_\RP(\Sf))$
(since \code{\RP.repeat} makes a call to \code{*$f$} and itself), we prove:\\
\mbox{}\hfill$
P_\RP \lectx \spcmod{S}{(A_\RP,\munit)}{H_\RP(\Sf)}
$.\hfill\mbox{}\\
Also, we verify $\SC$. For any $S$, we prove:\\
\mbox{}\hfill$
P_\SC \lectx \spcmod{S}{(A_\SC,\munit)}{S_\SC}
$.\hfill\mbox{}\\
Also, we verify $\AD$. For any $\Sf \sqsupseteq S_\SC$ (since \code{\SC.succ} is passed to \code{\RP.repeat})
and any $S \sqsupseteq H_\RP(\Sf)$ (since \code{\AD.add} makes a call to \code{\RP.repeat}), we prove:\\
\mbox{}\hfill$
P_\AD \lectx \spcmod{S}{(A_\AD,\munit)}{S_\AD}
$.\hfill\mbox{}\\
Finally, we instantiate the CRs with $\Sf = S_\SC$ and $S = H_\RP(S_\SC) \cup S_\SC\cup S_\AD$ and apply ACT to them with $\munit$ to \code{main()}:
\[
\small
\begin{array}{@{\!}l@{}l@{\ \ }l@{}}
  &
  P_\RP \link P_\SC \link P_\AD
  \\
  \lectx&
  \spcmod{S}{(A_\RP,\munit)}{H_\RP(S_\SC)} \!\link\! \spcmod{S}{(A_\SC,\munit)}{S_\SC} \!\link\! \spcmod{S}{(A_\AD,\munit)}{S_\AD}
  &
  \\
  \lectx&
  \toabs{A_\RP} \link \toabs{A_\SC} \link \toabs{A_\AD}
  &
\end{array}
\]

\hide{
\[
\begin{array}{@{}l@{~}l@{\ \ }l@{}}
  &
  P_\RP \link P_\SC \link P_\AD
  \\
  \lectx&
  \spcmod{S}{(A_\RP,\munit)}{H_\RP(S_\SC)} \!\link\! \spcmod{S}{(A_\SC,\munit)}{S_\SC} \!\link\! \spcmod{S}{(A_\AD,\munit)}{S_\AD}
  & \text{(by compositionality of CR)}
  \\
  \lectx&
  \toabs{A_\RP}{\setofz{\RP,\SC,\AD}} \link \toabs{A_\SC}{\setofz{\RP,\SC,\AD}} \link \toabs{A_\AD}{\setofz{\RP,\SC,\AD}}
  & \text{(by spec erasure thm.)}
\end{array}
\]}

As an advanced example, we also verify Landin's knot~\cite{irislecture} (see our Coq development~\cite{supplementary}).
We believe this pattern is a general solution to higher-order reasoning in practice
since it applies to all practical examples we can think of.




\section{Evaluation, related and future work} 
\label{sec:related}

\paragraph{Evaluation}
Our development comprises 37,329 SLOC of Coq (counted by \code{coqwc}), including 10,100 SLOC for all the examples in the paper and technical report~\cite{supplementary}.
For differential testing, we ran each example in two ways (by extracting both implementation and abstraction to OCaml) and compared the results. 
\originaltxt{For testing, we extracted both implementations and abstractions of all the examples in the paper and technical report~\cite{supplementary} to OCaml programs, executed them and compared the results.}%
Interestingly, we found two mis-downcast bugs in the \Echo{} example~\cite[\S3.4]{supplementary} by testing it before verification.


\paragraph{Related work}
Main related works are discussed in \Cref{sec:introduction}.
Here we discuss other related works.


Relational Hoare/separation logics~\cite{benton:relhoare,yang:relsep} establish refinement (or equivalence) between programs.
However, they do not support conditions on external functions but only on input and output states.
More specifically, they do not allow invoking unknown functions relying on their specifications.
Other logics~\cite{crellvm,steve:hoareitree} proving a form of refinement also only establish unconditional refinement.




There are \emph{non-relational} program logics (\ie proving safety not refinement or equivalence) that support higher-order specifications without using step-indexing.
First, the key idea of CFML~\cite{arthur:higher-imp, arthur:higher, arthur:pearl} to avoid step-indexing is essentially similar to our idea presented in \Cref{sec:advanced-higher},
although applied in a different setting (\ie unary program logic instead of conditional refinement).
Second, XCAP~\cite{xcap} and Bedrock~\cite{adam:bedrock13} use a syntactic technique to avoid step-indexing,
where higher-order predicates are treated as syntactic objects.

\hide{
There is another line of work~\cite{yang:relsep,benton:relhoare,crellvm} in relational separation logic that proves refinement (or
equivalence) when the whole program is verified in the logic. However, none of them support
vertical compositionality of each module as we do. Note that the intermediate specifications
used in \mwmidcond and \appmidcond does not match with the top-level specifications, which in
turn means there is no trivial way to emulate our reasoning in theirs.
}

\hide{
Thanks to our simple coinductive definition
of behavior, we have had no hassle in supporting termination-sensitive (\ie taking infinite
behaviors into account) refinement and did not discuss about it much.  However, this topic
itself is often considered worth an article and there are recent works~\cite{spies:transfinite, steve:hoareitree}
tackling this problem. These works
only considered a simple, closed setting and are irrelevant from our main contribution.
}

\hide{ There are variants of program logics~\cite{TODO-CFML,
    http://www.chargueraud.org/research/2020/seq_seplogic/seq_seplogic.pdf, TODO-Characteristic-Imperative-Programs} that support higher-order
  specifications by utilizing a higher-order quantification in the base logic (\eg
  Coq). However, they have different setting from ours that they pass first-class functions
  as arguments, not function pointers.  }

\hide{
There are variants of program logics~\cite{arthur:higher-imp, arthur:higher, arthur:pearl}
that support higher-order specifications but in a different setting from ours that they use
first-class functions, not function pointers.  A closer work in a setting similar to ours is
XCAP~\cite{xcap}. However, they have a serious limitation~\cite{indexed-xcap} that they only
support continuation-passing style program (\eg assembly) and does not scale well to
high-level languages (\eg C).  Another difference is that they needed a special extension
(impredicative polymorphism) which, in our approach, was not necessary.

}

\paragraph{Future work}
Since CCR is a new framework that spans refinement-style verification, logic-style verification, and testing,
there are various future research directions:
$(i)$ supporting (relaxed-memory) concurrency in the style of \iris{}~\cite{irisgroundup},
$(ii)$ embedding assembly into \ems{} in the style of CompCertM~\cite{CompCertM}, enabling verification of software composed of C and assembly,
which can be lowered to assembly via compilation with CompCertM, and
$(iii)$ developing property-based testing tools for efficient differential testing between an implementation and its abstraction.

\hide{
There have been many works in proving safety and refinement of programs in various directions but occasionally with certain restrictions.
Abstraction logic can be seen as a unifying theory, based on elementary mechanisms, that can subsume most of the works without such restrictions.
We will discuss those works and their restrictions if any.



\paragraph{Specifications as programs}

Refinement calculus~\cite{back2012refinement} understands Hoare-style specifications as programs and
provides refinement between them, which enjoys fully compositionality as in CR.
\cite{jeremie:lics, jeremie:thesis} recently made advances in this line of research,
where they also employ dual-nondeterminism and algebraic effects (similar to interaction trees but without extraction) like \crems.
However, unlike \crems, they do not support SL-style specifications.
\hide{
\paragraph{Dual nondeterminism} The concept is already present in the literature\cite{
tyrrell2006lattice}, but few of
them used it for writing specifications, and none connected it with SL. The one that is
closest to our approach is that of \cite{jeremie:lics, jeremie:thesis}, where dual
non-determinism is formalized using domain theory, used in writing specifications, and such
specifications are verified with \yj{TODO}.  Also, they employ algebraic effects which
corresponds to interaction trees, but they do not support extraction to OCaml. \yj{COMPLETELY REWRITE THIS PARAGRAPH}
}




\paragraph{Separation Logics}

First of all, compared to the state-of-the-art separation logics such as Iris and VST, \CREMS does not support concurrency yet
although we plan to extend \crems to support it following their approaches.


There have been works based on separation logics that go beyond safety such as CaReSL\cite{caresl} and ReLoC\cite{reloc}.
Like \crems, they establish contextual refinement using SL but in a restricted setting that
does not allow transformations essentially relying on logical specifications of external modules.

Using \iris, \cite{sandbox} establishes guarantee of desired
properties on observable traces (\ie a sequence of system calls),
instead of safety guarantee, in the presence of unverified contexts,
but in a restricted setting that does not allow the contexts to invoke system calls.


\hide{
As discussed (\Cref{sec:introduction}), standard version of SL
has several limitations and there are various works each extending with different directions.


Notable works that go beyond mere safety are CaReSL\cite{caresl} and ReLoC\cite{reloc}
but they have different focus with us. On one hand, they have advantages of
supporting concurrency and general higher-order features, utilizing step-index. On the other
hand, while they do establish CR, they also inherit the limitations of it
(\Cref{sec:introduction}) that they cannot use SL specifications for shared
states. Consequently, their verification examples are restricted to abstracting non-shared
states (references that have not been leaked to outside). Nonetheless, such abstraction is
not sound if an arbitrary context illegally accesses such location, so they rule out such
context by typing.
}





\paragraph{Contextual Refinement}

Certified Abstraction Layers (CAL)\cite{gu:dscal,ccal} proved effectiveness of contextual refinement in large scale verification
by verifying a realistic operating system.
Compared to \crems, although it supports concurrency, CAL is limited in a few aspects.
For example, CAL does not support SL-style specifications and thus does not allow implementations to use shared resources across modules.
Also it does not allow mutual recursion between modules.

}




\hide{

We present a comprehensive theory combining the benefits of contextual refinement and separation logic, together with practical tools,
using the key idea of \kwchoose{} and \kwtake{} that gives an illusion of passing any information to anyone without involving physical operations.
As future works, we plan to extend \CREMS to support concurrency,
and also develop testing tools that can efficiently find bugs that breaks desired contextual refinement,
which may also give a certain level of confidence without verification.

}

\hide{
\myparagraph{Future Work} We are interested in (i) extending the framework to concurrent
setting, and (ii) supporting automated testing seriously. For the former, we believe that our
most fundamental ideas will scale to concurrent setting. Furthermore, it would be interesting to
adopt the ideas from \cite{liang:pp} to specify and verify progress properties with CR. For
the latter, we are interested in adopting SMT-solver\cite{???} or property-based
testing\cite{???}. We believe that gradual abstraction will play an important role in doing
so, as noticed in \cite{ARMADA}.
}

\newpage
\balance
\bibliography{references}



\end{document}